\documentclass{aa}  

\usepackage{graphicx}
\usepackage{txfonts}

\usepackage{graphicx}   
\usepackage{amsmath}    
\usepackage{amssymb}    
\usepackage{enumerate}
\usepackage{longtable}
\usepackage{caption}
\usepackage{textcomp}

\begin{document} 

    \title{HOLISMOKES. VIII. High-redshift, strong-lens search in the Hyper Suprime-Cam Subaru Strategic Program\thanks{Tables~\ref{tb:gradeA} and \ref{tb:gradeB} are only available in electronic form
at the CDS via anonymous ftp to cdsarc.u-strasbg.fr (130.79.128.5)
or via http://cdsarc.u-strasbg.fr/viz-bin/cat/J/A+A/662/A4.}}

    \author{Yiping Shu\inst{1, 2} 
    \and Raoul Ca{\~n}ameras\inst{1} 
    \and Stefan Schuldt\inst{1, 3} 
    \and Sherry H. Suyu\inst{1, 3, 4} 
    \and Stefan Taubenberger\inst{1}
    \and Kaiki Taro Inoue\inst{5}
    \and Anton T. Jaelani\inst{6}
    }

    \institute{Max-Planck-Institut f\"{u}r Astrophysik, Karl-Schwarzschild-Str. 1, 85748 Garching, Germany\\
    \email{ypshu@mpa-garching.mpg.de}
    \and
    Ruhr University Bochum, Faculty of Physics and Astronomy, Astronomical Institute (AIRUB), German Centre for Cosmological Lensing, 44780 Bochum, Germany
    \and
    Technische Universit\"{a}t M\"{u}nchen, Physik Department, James-Franck Str. 1, 85748 Garching, Germany
    \and 
    Institute of Astronomy and Astrophysics, Academia Sinica, 11F of ASMAB, No.1, Section 4, Roosevelt Road, Taipei 10617, Taiwan
    \and
    Faculty of Science and Engineering, Kindai University, Higashi-Osaka, 577-8502, Japan
    \and
    Astronomy Research Group and Bosscha Observatory, FMIPA, Institut Teknologi Bandung, Jl. Ganesha 10, Bandung 40132, Indonesia
    }

   \date{Received xxx; accepted xxx}

 
  \abstract{We carry out a search for strong-lens systems containing high-redshift lens galaxies with the goal of extending strong-lensing-assisted galaxy evolutionary studies to earlier cosmic time. Two strong-lens classifiers are constructed from a deep residual network and trained with datasets of different lens-redshift and brightness distributions. We classify a sample of 5,356,628 pre-selected objects from the Wide-layer fields in the second public data release of the Hyper Suprime-Cam Subaru Strategic Program (HSC-SSP) by applying the two classifiers to their HSC $gri$-filter cutouts. Cutting off at thresholds that correspond to a false positive rate of $10^{-3}$ on our test set, the two classifiers identify 5,468 and 6,119 strong-lens candidates. Visually inspecting the cutouts of those candidates results in 735 grade-A or B strong-lens candidates in total, of which 277 candidates are discovered for the first time. This is the single largest set of galaxy-scale strong-lens candidates discovered with HSC data to date, and nearly half of it (331/735) contains lens galaxies with photometric redshifts above 0.6. Our discoveries will serve as a valuable target list for ongoing and scheduled spectroscopic surveys such as the Dark Energy Spectroscopic Instrument, the Subaru Prime Focus Spectrograph project, and the Maunakea Spectroscopic Explorer.}

   \keywords{Galaxies: evolution -- Gravitational lensing: strong -- Methods: data analysis
    }

   \titlerunning{HOLISMOKES VIII.}
   \maketitle
%

\section{Introduction}

The strong gravitational lensing effect is a powerful and robust mass probe that can deliver precise and accurate measurements of the total mass (including dark matter) in the central regions of galaxies at extragalactic distances. Studies of strong-lens systems have successfully measured dark matter and stellar mass distributions and their evolution in distant galaxies, which have deepened our understanding of galaxy formation and evolution \citep[e.g.][]{Treu06, Koopmans06, Auger10, Bolton12a, Brewer14, Shu15, Shu16a}. Detections of dark-matter substructures beyond the local Universe and measurements of their masses from strong lensing have placed constraints on the sub-halo mass function and the nature of dark matter \citep[e.g.][]{Vegetti10, Vegetti12, Fadely12, Nierenberg14, Hezaveh16, Inoue16}. In addition, the lensing magnification effect can be exploited to study high-redshift objects in detail by overcoming the sensitivity and/or resolution limitations of current facilities \citep[e.g.][]{Christensen12, Bussmann13, Stark15, Shu16b, Marques-Chaves17, Marques-Chaves18, Marques-Chaves20, Shu21}. Moreover, strongly lensed variable sources, such as quasars and supernovae (SNe), have evolved into an independent and compelling cosmological probe \citep[e.g.,][]{Suyu10a, Suyu13, Suyu17, Grillo18, Wong20, Millon20}, which is one of the main motivations for our Highly Optimised Lensing Investigations of Supernovae, Microlensing Objects, and Kinematics of Ellipticals and Spirals (HOLISMOKES) programme \citep{Suyu20}. 

Various techniques have been developed to identify the intrinsically rare strong-lens systems. The most productive ones to date are imaging-based methods, which have discovered $\approx 400$ confirmed strong-lens systems\footnote{http://admin.masterlens.org/index.php}$^{,}$\footnote{http://www-utap.phys.s.u-tokyo.ac.jp/\url{~}oguri/sugohi/}$^{,}$\footnote{ https://research.ast.cam.ac.uk/lensedquasars/index.html}\citep[e.g.][]{Browne03, More12, Stark13, sugohi_I, Lemon18, Shu18, Shu19, sugohi_IV, Desira22}. In this work, we consider a strong-lens system as confirmed if multiple lensed images are detected and the lens and source redshifts are spectroscopically measured. Over the past two decades, spectroscopy-based methods have heavily exploited large-scale spectroscopic surveys and discovered more than 200 confirmed strong-lens systems \citep[e.g.][]{Bolton04, Bolton08, Treu11, Brownstein12, Courbin12, Shu16b, Shu16a, Oldham17, Shu17}. Very recently, variability-based methods, which are particularly useful for discovering strongly lensed variable sources, have gained momentum and will undoubtedly play a crucial role in the ongoing and upcoming time-domain surveys \citep[e.g.][]{Kostrzewa-Rutkowska18, Chao20, Chao20b, YShu21, Bag21}.  

Although the total number of confirmed strong-lens systems have reached $\approx 600^{1}$, many scientific applications call for more systems and a more thorough coverage of the phase space. For example, a lot of effort has been made to search for strongly lensed SNe, which is expected to provide tighter constraints on the Hubble constant compared with strongly lensed quasars \citep[e.g.][]{Oguri03, Goldstein17, Wojtak19, Huber21b, Huber21, Bayer21, Ding21}. Two efficient approaches of catching such rare and short-lived lensing events are (1) cross-matching transient alerts from time-domain surveys with known strong-lens systems, and (2) carrying out dedicated monitorings of known strong-lens systems with high expected lensed SN rates \citep[e.g.][]{Shu18a, Ryczanowski20, Craig21}. Both of these approaches benefit greatly from discovering more strong-lens systems. Additionally, strong-lensing-assisted evolutionary analyses have so far been limited to low- and intermediate-redshift galaxies due to the lack of galaxy-galaxy strong-lens systems with high-redshift lens galaxies. Among all confirmed galaxy-galaxy strong-lens systems, only a handful contain lens galaxies at redshifts beyond 0.8 \citep[e.g.][]{Wong14, Canameras17}. On the other hand, high-redshift galaxies are crucial to understanding galaxy evolution as they are expected to undergo more frequent and vigorous transitions. Recently, the combination of wide-field imaging surveys and machine learning algorithms has led to a big leap in strong lens discoveries. A few thousand new strong-lens candidates have been uncovered by classifiers built upon supervised or unsupervised algorithms \citep[e.g.][]{Jacobs19b, Petrillo19, Canameras20, Canameras21, Huang20, Li20, Huang21, Li21, Stein21, Rojas21, Savary21}. Future surveys, such as the Rubin Observatory Legacy Survey of Space and Time \citep[LSST,][]{Ivezic19}, Euclid \citep{Euclid}, and the Chinese Space Station Optical Survey \citep[CSS-OS,][]{Zhan18}, expect to deliver $\sim 10^5$ strong-lens systems \citep[e.g.][]{Collett15}.

In this work, we focused on extending strong-lensing-assisted evolutionary analyses to earlier cosmic time by searching for high-redshift strong lenses in the Wide-layer data from the second public data release (PDR2) of the Hyper Suprime-Cam Subaru Strategic Program \citep[HSC-SSP,][]{Aihara2019}. In Section~\ref{sect:data}, we describe the HSC-SSP PDR2 data and define our parent sample. Section~\ref{sect:method} explains the construction and training of our two strong-lens classifiers based on a deep residual network, and the performance of the two classifiers is shown in Section~\ref{sect:performance}. Discovered strong-lens candidates are presented in Section~\ref{sect:results}. Six candidates that show two sets of spectral features at different redshifts in auxiliary spectroscopic data are reported in Section~\ref{sect:notes}. Discussions and conclusions are provided in Sections~\ref{sect:discussion} and \ref{sect:conclusion}. To compute the Einstein radii, we adopt a flat $\Lambda$CDM cosmology with $\Omega_{\rm m} = 1 - \Omega_{\rm \Lambda} = 0.32$ \citep{Planck20} and $H_0 =$ 72 km s$^{-1}$ Mpc$^{-1}$ \citep{Bonvin17}.

\section{Data}
\label{sect:data}

In HSC-SSP PDR2, the Wide-layer data cover $\approx$300 deg$^2$ to the nominal depths in all five filters (i.e. $grizy$) and additional $\approx$1,100 deg$^2$ in at least one filter and one exposure. For the PDR2 Wide layer, the median 5$\sigma$ depths (for point sources) in $grizy$ filters are 26.6, 26.2, 26.2, 25.3, and 24.5 mag and the median seeings in $grizy$ filters are 0\farcs77, 0\farcs76, 0\farcs58, 0\farcs68, 0\farcs68, respectively. A full overview of HSC-SSP PDR2 can be found in \citet{Aihara2019}. 
For our high-redshift strong-lens search, we selected objects that are extended and likely located at high redshifts based on their $g-r$ and $g-i$ colours. To be more specific, we selected objects in the PDR2 Wide layer, that is the {\tt pdr2\_wide.forced} table, that satisfy the following criteria:
\begin{enumerate}
    \item {\tt isprimary} is {\tt True}
    \item {\tt i\_extendedness\_value=1}
    \item {\tt [grizy]\_sdsscentroid\_flag} is {\tt False}
    \item {\tt [grizy]\_pixelflags\_edge} is {\tt False}
    \item {\tt [grizy]\_pixelflags\_interpolatedcenter} is {\tt False}
    \item {\tt [grizy]\_pixelflags\_saturatedcenter} is {\tt False}
    \item {\tt [grizy]\_pixelflags\_crcenter} is {\tt False}
    \item {\tt [grizy]\_pixelflags\_bad} is {\tt False}
    \item {\tt [grizy]\_cmodel\_flag} is {\tt False}
    \item {\tt g\_cmodel\_mag} < 26.0
    \item {\tt r\_cmodel\_mag} < 26.0
    \item {\tt i\_cmodel\_mag} < 26.0
    \item 0.6 < {\tt g\_cmodel\_mag-r\_cmodel\_mag} < 3.0
    \item 2.0 < {\tt g\_cmodel\_mag-i\_cmodel\_mag} < 5.0
\end{enumerate}
This query returns 5,356,628 unique HSC objects in total, which form the parent sample of this lens search project. Here, criteria 3--12 are used to remove objects with unreliable photometry \citep[e.g.][]{Tanaka2018, Schuldt21b}, and the colour-colour cuts in criteria 13--14 are directly taken from \citet{Jacobs19a} to select red and potentially high-redshift galaxies. The HSC CModel photometry algorithm is presented in detail in \citet{Bosch18}. In summary, the single-filter imaging data of an object are fitted separately with an elliptical exponential model or with an elliptical de Vaucouleurs model, where each model is convolved with the point spread function (PSF). The CModel magnitude is subsequently computed from a composite model that is constructed as a linear combination of the previous exponential and de Vaucouleurs models, which best fit the imaging data. Since the CModel photometry is based on a reasonable analytical description of galaxy morphology, we expect it to provide more robust colour estimates than the fixed-aperture or Kron photometry that are also available in PDR2, especially for lens galaxies in strong-lens systems. We find that criteria 13--14 manage to substantially reduce the sample size and at the same time maintain a high completeness rate for high-redshift lens galaxies. Removing criteria 13--14 in the above query would have resulted in a sample of 79,577,619 unique extended objects, which in turn would have posed challenges to not only the final lens search but also the initial imaging data retrieval. On the other hand, \citet{Jacobs19a} simulated 10,000 $z > 0.8$ elliptical galaxies with lensing features superimposed and found that $\gtrsim 90\%$ of the simulated lenses can be recovered with these two colour-colour cuts. In addition, we examined the colour distributions of strong-lens candidates discovered in the HSC footprint by the Survey of Gravitationally-lensed Objects in HSC Imaging (SuGOHI) project \citep{sugohi_I, sugohi_II, sugohi_IV, sugohi_V, sugohi_VI}. Every SuGOHI strong-lens candidate is assigned a grade of A (definite), B (probable), or C (possible) and a lens type from GG (galaxy-galaxy), GQ (galaxy-quasar), CG (cluster- or group-galaxy), or CQ (cluster- or group-quasar). As we are particularly interested in galaxy-galaxy strong lenses, we focused on the 99 SuGOHI grade-A or B GG strong-lens candidates that have lens galaxies fulfilling criteria 1--12. The lens galaxies in those strong-lens candidates are primarily luminous red galaxies selected according to the criteria defined in \citet{Dawson13}. They span a wide redshift range from 0.2 to 1.0\footnote{According to the spectroscopic or photometric redshifts available on http://www-utap.phys.s.u-tokyo.ac.jp/\url{~}oguri/sugohi/}. We note that candidates from \citet{sugohi_VI} are not considered here because some GG strong-lens candidates therein are actually cluster- or group-scale lenses. Among the selected 99 SuGOHI strong-lens candidates, 92 (or $\approx 93\%$) further pass the colour-colour cuts in criteria 13--14. Limiting to the selected SuGOHI candidates with lens galaxy (spectroscopic or photometric) redshifts above 0.8, 4/5 (or $80\%$) pass the colour-colour cuts. Although the colour-colour cuts were originally defined in the photometric system of the Dark Energy Survey, we expect them to be similarly effective in the HSC photometric system given the minor difference between them \citep{Abbott21} and the encouraging results from the SuGOHI sample.

The HSC $gri$-filter cutouts (72 pixel $\times$ 72 pixel, 1 pixel = 0\farcs17) centred on the 5,356,628 objects in our parent sample are retrieved from the PDR2 image cutout service. Photometry \citep[CModel magnitudes from the {\tt pdr2\_wide.forced} table,][]{Aihara2019} and photometric redshift \citep[{\tt photoz\_best} from the {\tt pdr2\_wide.photoz\_mizuki} table,][]{Tanaka2018} for every object in the parent sample are also retrieved from the HSC CAS Search service. The parent sample covers roughly 960 deg$^2$. 

\section{Strong-lens classifier construction}
\label{sect:method}

We constructed our strong-lens classifiers based on the deep residual network, {\tt deeplens\_classifier}, pre-built in the {\tt CMU DeepLens} package \citep{Lanusse18}. Deep residual networks (resnets), a variation of convolutional neural networks, have become the current state-of-the-art imaging recognition algorithm, and {\tt CMU DeepLens} adopts a specific resnet architecture proposed by \citet{He16}. Among the nine different lens-finding methods in the strong gravitational lens finding challenge \citep{Metcalf19}, {\tt CMU DeepLens} delivered the highest area under the receiver operating characteristic curve (AUROC) value, which is the most commonly used evaluation metric for classification problems. It is also top-ranked on TPR$_0$ and TPR$_{10}$, which correspond to the highest true positive rate reached before more than 0 and 10 false positives occur, respectively. We therefore chose {\tt deeplens\_classifier} from {\tt CMU DeepLens} as our baseline model, and a full description of the network architecture can be found in \citet{Lanusse18}. The {\tt deeplens\_classifier} network is constructed such that it returns a number from 0 to 1 for every input system, which is referred to as the network score $p_{\rm resnet}$ in this work. 

The {\tt deeplens\_classifier} network takes several parameters that determine how the actual training is done. In particular, {\tt learning\_rate} sets the initial learning rate, {\tt learning\_rate\_steps} sets the number of learning rate updates during training, {\tt learning\_rate\_drop} sets the amount by which the learning rate is updated, and {\tt n\_epochs} sets the total number of training epochs. For example, the network that delivered the highest AUROC value in the strong gravitational lens finding challenge had {\tt learning\_rate}$=0.001$, {\tt learning\_rate\_steps}$=3$, {\tt learning\_rate\_drop}$=0.1$, and {\tt n\_epochs}$=120$, which correspond to a starting learning rate of $0.001$ that is multiplied by 0.1 every 40 epochs. We always use a {\tt learning\_rate\_drop} of $0.1$ for our classifiers. 

In this work, we test two strong-lens classifiers. The main difference between the two is the properties of mock lenses in the training set. This allows us to investigate, among others, the impact of the training set on classifier performance. In addition, combining the results from the two classifiers yields a much more complete sample of strong-lens candidates, as we demonstrate later.

\subsection{Classifier-1}

\subsubsection{Training and validation datasets}
\label{sect:TraningSet}

As the sample size of confirmed strong lenses is still small (of the order of $10^3$), mock lens systems need to be created for training and validation. We tried to be as realistic as possible by using observed data of real galaxies to make the mock systems. Following \citet[][C21 hereafter]{Canameras21}, we selected $\approx$80,000 galaxies from data release 14 of the Sloan Digital Sky Survey \citep[SDSS,][]{DR14} that are also in the HSC footprint and have measured spectroscopic redshifts and velocity dispersions \citep{Bolton12} as the lens sample. We directly took HSC $gri$-filter cutouts (72 pixel $\times$ 72 pixel) centred on those lens galaxies as the base layer. As a result, mock lens systems naturally include various observational effects, such as galaxy colour gradients, seeing variations, neighbouring and line-of-sight contaminants, and artefacts, that are also present in the parent sample. To further enlarge the lens sample, we rotated every galaxy in the lens sample by 90\textdegree, 180\textdegree, and 270\textdegree, and considered them as different lens galaxies. This implies that each galaxy in the lens sample is used four times at most. For the source sample, we used $\approx$1,200 high signal-to-noise ratio (S/N) galaxies in the Hubble Ultra Deep Field with secure spectroscopic redshifts \citep{Inami17}. We converted images of the selected source galaxies in HST bands (F435W, F606W, and F775W) to HSC $gri$ filters using the method in \citet{Canameras21}. 

\begin{figure*}
    \centering
    \includegraphics[width=0.96\textwidth]{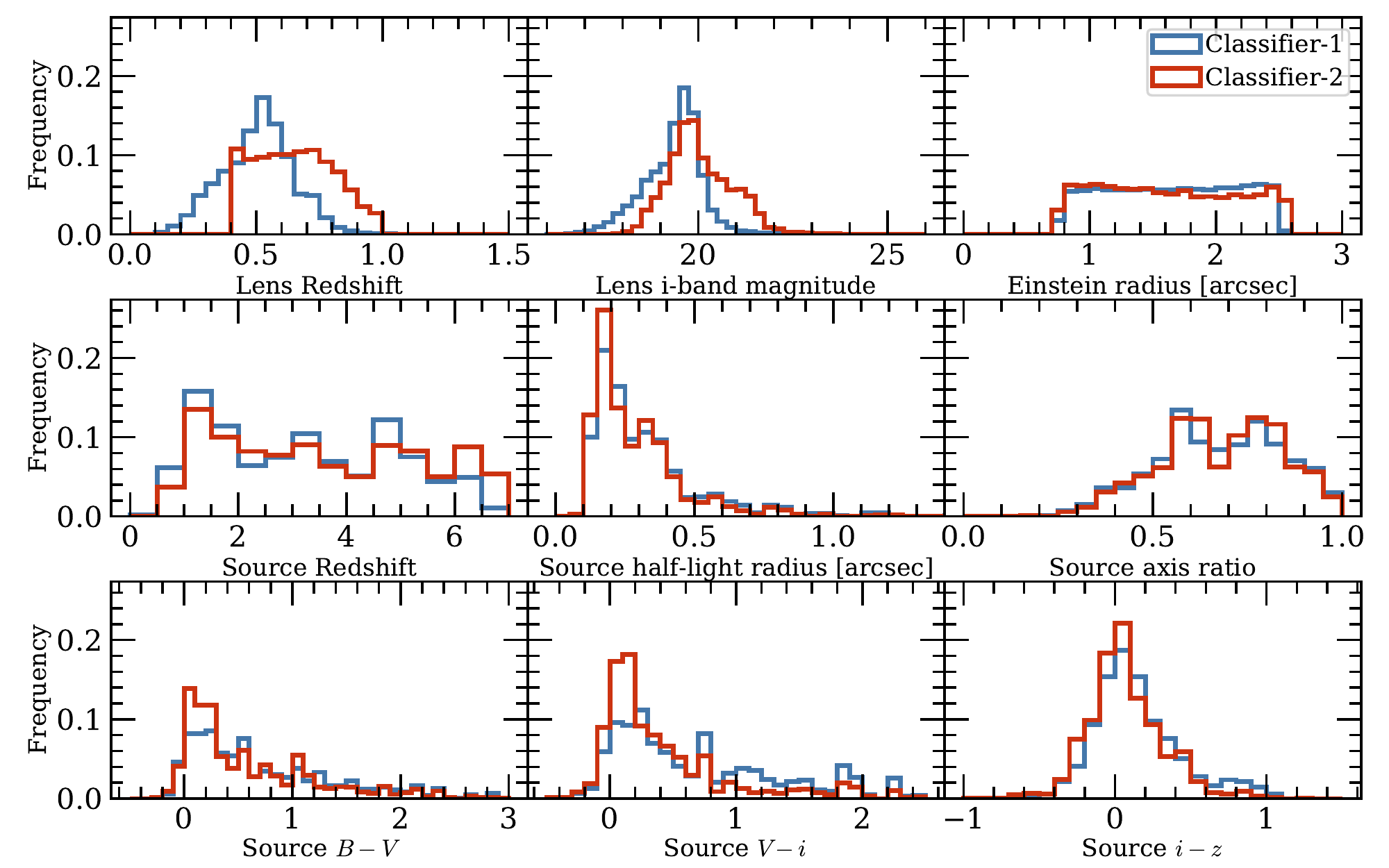}
    \caption{Distributions of lens galaxy redshift, lens galaxy $i-$band magnitude, Einstein radius, source galaxy redshift, source galaxy half-light radius, source galaxy axis ratio, and source galaxy $B-V$, $V-i$, and $i-z$ colours for mock lenses in the training sets for Classifier-1 (blue) and Classifier-2 (red).}
    \label{fig:TrainingSet_props}
\end{figure*}
\begin{figure*}
    \centering
    \includegraphics[width=0.47\textwidth]{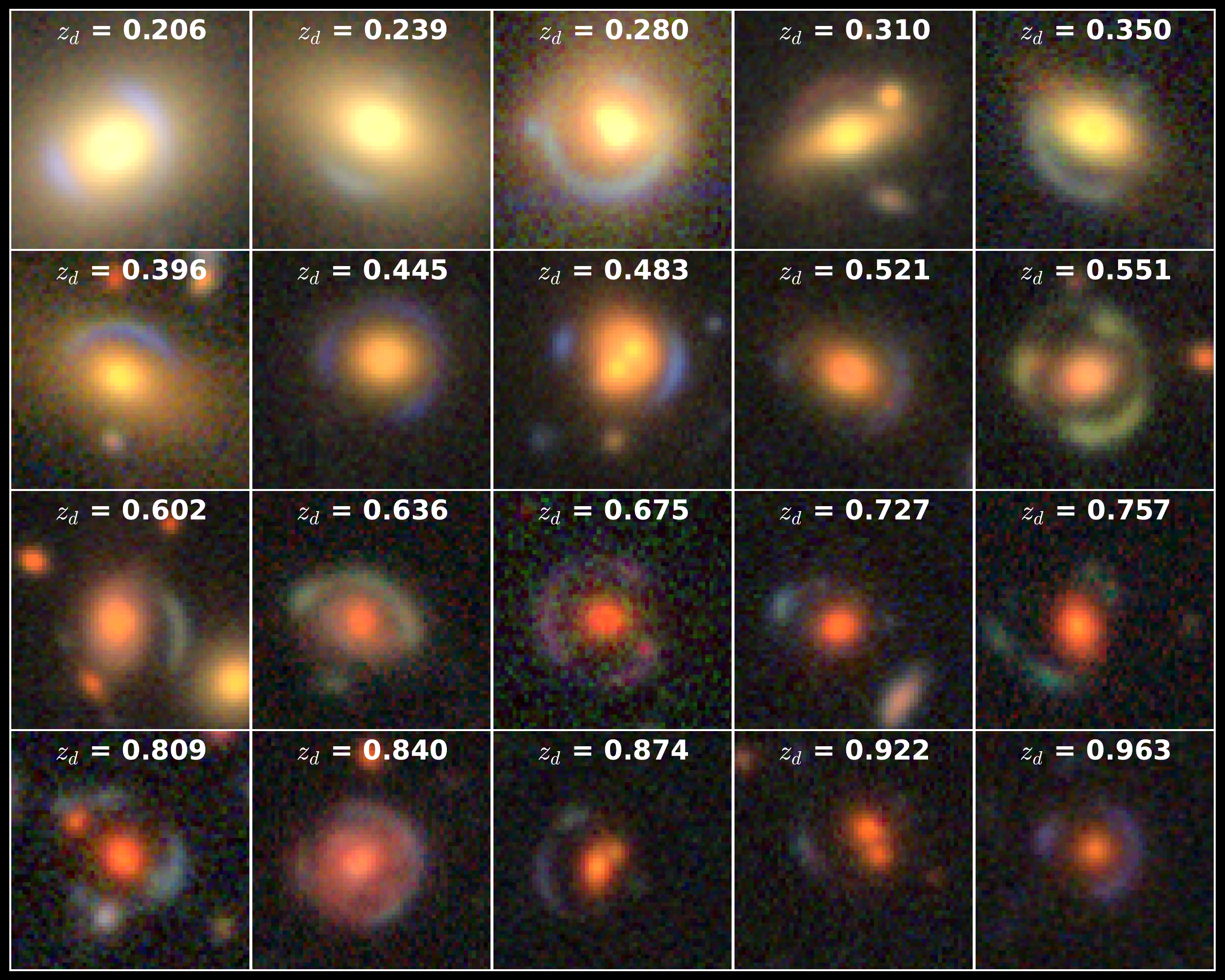}
    \includegraphics[width=0.47\textwidth]{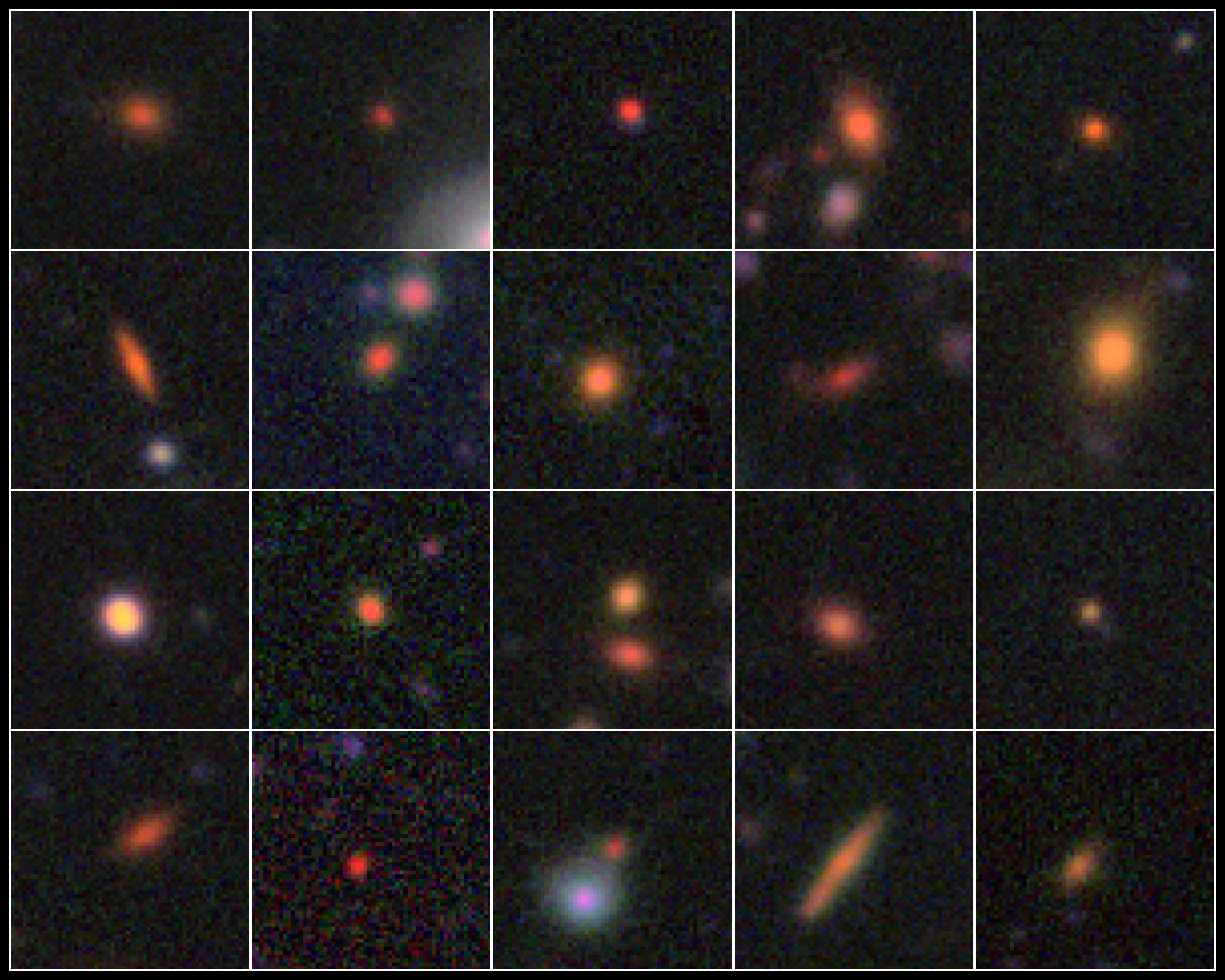}
    \caption{Colour composite images of 20 mock lenses (left, ordered by lens galaxy redshift) and 20 non-lens examples (right) selected from the training set for Classifier-1.}
    \label{fig:TrainingSet}
\end{figure*}

Similarly to procedures used in \citet{Canameras20, Canameras21} and \citet{Schuldt21a}, we modelled the effective lensing potential as two components: a projected lens mass component characterised by a singular isothermal ellipsoid (SIE) profile and an external shear. The axis ratio and position angle of the SIE profiles were set to values inferred from the lens surface-brightness distribution in the HSC $i$ band. The external shear strength was randomly drawn from a Gaussian distribution with mean 0 and standard deviation 0.058 \citep[e.g.][]{Wong11, Faure11} and the position angle was randomly chosen from 0\textdegree \ to 180\textdegree. For every lens galaxy, we randomly paired it with a galaxy from the source sample that is at a redshift higher than the lens galaxy. The Einstein radius of the SIE profile can then be computed from the lens and source redshifts and the lens velocity dispersion. The selected source galaxy is randomly placed with a requirement that its centroid needs to be at a location with a total magnification of 5 or more. We used {\tt GLEE} \citep[][]{Suyu10, Suyu12} to generate the lensed image of the source, which is further downsampled to the HSC pixel size and convolved with the PSF at the location of the lens provided by the HSC PSF picker. We required the brightest pixel in the lensed image to be brighter than the corresponding pixel in the base layer in either $g$- or $i$-band. Otherwise, we draw a new source position, generate the lensed image, and compare. This process can be iterated 40 times at most, after which point the brightness of the selected source galaxy is boosted by 0.5 mag in all three bands and the whole process is repeated. If the requirement is still not satisfied after boosting the selected source by 5 mag, a new source galaxy is selected from the source sample. Once the requirement is satisfied, the lensed image is added to the base layer to produce the composite image of a mock lens system. 

For this classifier, we specifically selected 43,500 mock lens systems that produce a close to uniform Einstein radius distribution between 0\farcs75 and 2\farcs5 as positive examples. The Einstein radius is the single most important quantity of a strong-lens system, and is determined primarily by the lens galaxy mass with an additional dependence on the lens and source redshifts. We choose a uniform Einstein radius distribution so that the classifier is equally sensitive to galaxy-scale strong-lens systems with different image separations. We tried training with mock lenses that have more naturally distributed Einstein radii, that is starting from 0\farcs75 and decreasing towards larger radii. The corresponding classifier had a lower overall TPR and failed to recover some of the obvious strong-lens candidates with large Einstein radii in the test set. To ensure the translation invariance of the classifier, for each mock lens system we extracted a 60 pixel $\times$ 60 pixel $gri$ cutout (roughly $10^{\prime \prime} \times 10^{\prime \prime}$) randomly centred within $\pm$ 5 pixels in both the R.A. and Decl. directions of the centre of the original cutout (72 pixel $\times$ 72 pixel), and we refer to the 43,500 cutouts as the lens dataset. Considering that the largest Einstein radii of our mocks are 2\farcs5 and shifts up to 0\farcs85 in each direction are applied, $10^{\prime \prime} \times 10^{\prime \prime}$ cutouts are needed and are sufficient to ensure all the lensing features are seen by the classifier. Using larger cutouts will presumably lead to classifier performance degradation as the chance of contamination due to irrelevant objects in the cutouts increases quadratically with the cutout size. As indicated by Figure~\ref{fig:TrainingSet_props}, the redshift distribution of lens galaxies in this training set peaks at $\approx0.55$. The $i$-band magnitude distribution of lens galaxies peaks at $\approx 19.5$ mag and drops rapidly towards the faint side. In fact, the magnitude distribution of the lens galaxies, which are all spectroscopically-observed galaxies in the SDSS surveys, is primarily due to SDSS selection effects. In SDSS-III, galaxies selected for spectroscopic observations are all brighter than $i = 19.9$ \citep{Dawson13}, and the faint limit for galaxy target selection extends to $i \leq 21.8$ in SDSS-IV \citep{Prakash16}. Distributions of several source galaxy properties are extracted from \citet{Beckwith05} and \citet{Inami17} and shown in Figure~\ref{fig:TrainingSet_props}. We note that the source redshift distribution is biased because of the applied artificial source brightness boosting (by up to 5 mag) during the generation of mocks.

To construct the non-lens examples for training and validation, we first randomly select 48,213 objects from the parent sample. To further clean this subset, we cross-matched them with a sample of 10,241 known strong lenses and strong-lens candidates (referred to as the known strong lens compilation hereafter) compiled from the literature \citep[e.g.][]{Diehl17, sugohi_I, sugohi_II, Petrillo19, Jacobs19a, Jacobs19b, sugohi_IV, sugohi_V, sugohi_VI, Huang20, Huang21, Canameras20, Li20, Canameras21, Li21, Rojas21, Savary21} using a matching radius of 30 arcsecs, and we removed the 114 matches. Considering the typical lensing rate of $10^{-4}$--$10^{-3}$ \citep[e.g.][]{Browne03, Bolton04, OM10, Treu10}, the remaining 48,099 objects are expected to be sufficiently pure. Among them, 43,500 objects are randomly selected as the final non-lens examples (to match the size of the lens dataset). Similarly, a random shift within $\pm 5$ pixels in both directions is applied simultaneously to the $gri$-filter cutouts of each non-lens example. The shifted $gri$-filter cutouts of the 43,500 objects are trimmed to 60 pixel $\times$ 60 pixel and form the non-lens dataset. 

The lens and non-lens datasets are merged into a single dataset, which is then randomly shuffled. 80\% of the shuffled dataset is used for training and the remaining 20\% is used for validation. Twenty mock lens systems and twenty non-lens systems randomly selected from the training set are shown in Figure~\ref{fig:TrainingSet} as an illustration. 

\subsubsection{Test dataset}
\label{sect:TestSet}

To construct the non-lens examples for the test set, we first randomly selected 53,570 objects from the parent sample. To further clean this subset, we cross-matched them with the known strong lens compilation from the previous step and the 43,500 non-lens examples used for training and validation using a matching radius of 30 arcsecs, and we removed the 152 and 1,649 matches. 50,000 objects were randomly selected from the remaining objects, and their $gri$-filter cutouts were trimmed to 60 pixel $\times$ 60 pixel and form the non-lens examples of the test set.

To construct the lens examples for the test set, we used strong lenses and strong-lens candidates from the SuGOHI project. The SuGOHI project has discovered 2,002 strong lenses and strong-lens candidates based on HSC imaging data \citep{sugohi_I, sugohi_II, sugohi_IV, sugohi_V, sugohi_VI}, of which 1,411 systems pass our selection criteria in Section~\ref{sect:data} and are 
included in our parent sample. As we are particularly interested in our network's ability to discover galaxy-galaxy strong lenses, we only included 23 grade-A and 69 grade-B galaxy-galaxy strong-lens candidates from the 1,411 SuGOHI systems in the
test set. Again, their $gri$-filter cutouts are trimmed to 60 pixel $\times$ 60 pixel and form the lens examples of the test set. For the sake of simplicity, candidates from \citet{sugohi_VI} are also not included in this step because some classified GG strong-lens candidates therein are actually cluster- or group-scale systems.

\subsubsection{Network tuning}

To quantify the network performance, we examined the true positive rate (TPR) and false positive rate (FPR). The TPR and FPR are defined as follows: 
\begin{equation}
    \text{TPR} = \frac{\text{Number of lenses that are correctly classified as lenses}}{\text{Number of lenses in a dataset}},
\end{equation}
\begin{equation}
    \text{FPR} = \frac{\text{Number of non-lenses that are mis-classified as lenses}}{\text{Number of non-lenses in a dataset}}.
\end{equation}
As mentioned previously, the network performance is usually measured by the AUROC metric for such a classification problem. The receiver operating characteristic (ROC) curve is the relation between TPR and FPR when the network score threshold varies from 0 to 1, and the AUROC is the integration of the ROC curve. For reference, a perfect classifier has an AUROC of 1.0, which is the best possible value, and a classifier that makes random predictions has an AUROC of 0.5. 

For this classifier, we explore three different options for network parameters {\tt learning\_rate}, {\tt learning\_rate\_steps}, and {\tt n\_epochs}. The first option corresponds to the default values that delivered the highest AUROC value in the Strong Gravitational Lens Finding Challenge, that is $[0.001, 3, 120]$ (in the format of [{\tt learning\_rate}, {\tt learning\_rate\_steps}, {\tt n\_epochs}]. The other two options are $[0.01, 4, 160]$ and $[0.1, 5, 200]$. The network that is trained with $[0.01, 4, 160]$ has the highest AUROC on the test dataset, and it was therefore chosen to be the final network for Classifier-1.


\begin{figure*}
    \centering
    \includegraphics[width=0.48\textwidth]{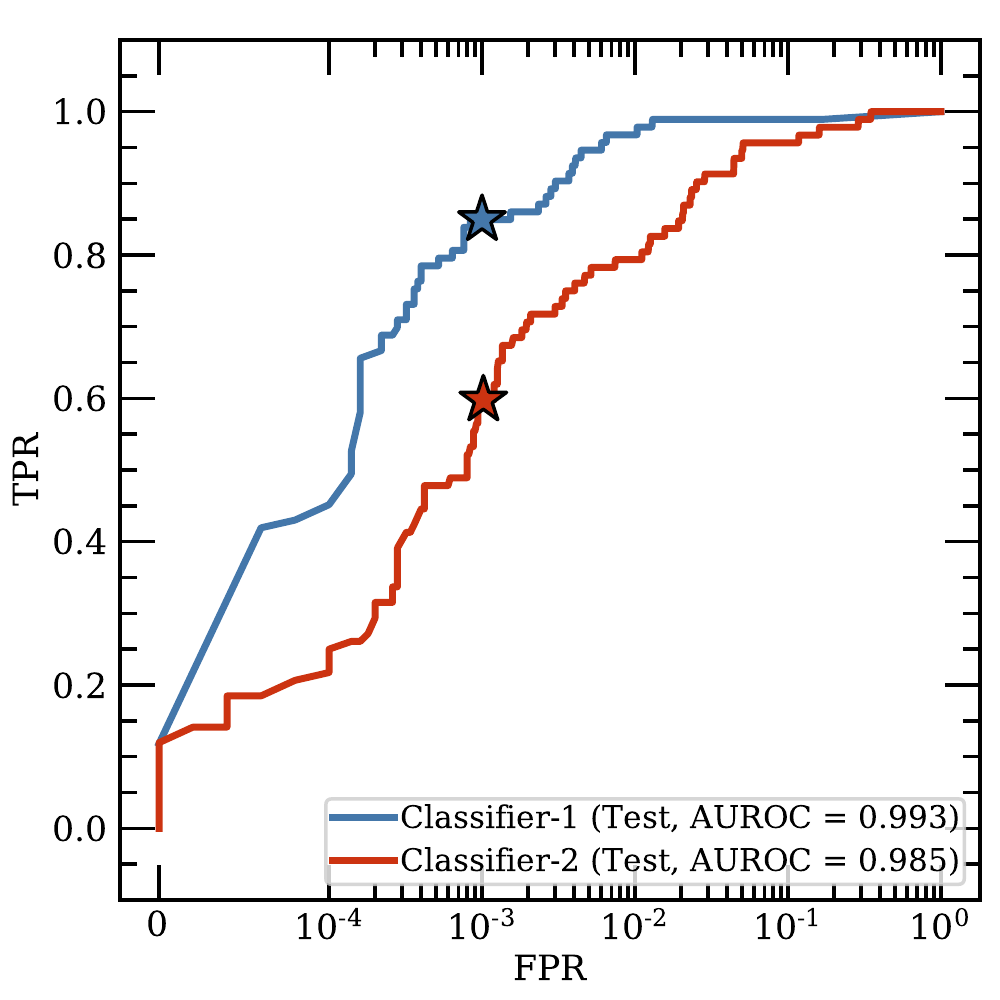}
    \includegraphics[width=0.48\textwidth]{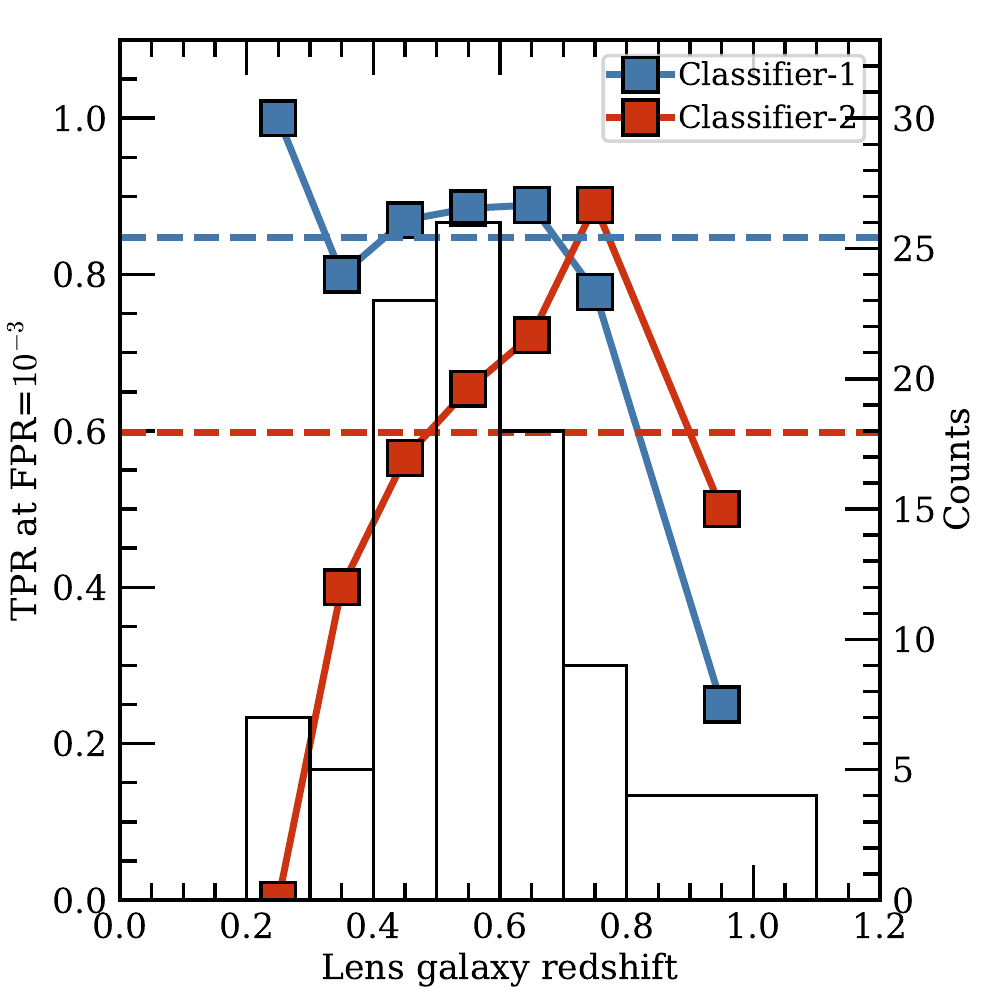}
    \caption{Performances of the two classifiers. \emph{Left:} ROC curves based on the test sets for Classifier-1 (blue) and Classifier-2 (red). The x-axis is scaled such that 0--$10^{-4}$ is in a linear scale and $10^{-4}$--1 is in a logarithmic scale. The two star symbols correspond to FPR$=10^{-3}$. \emph{Right:} TPR at an FPR of $10^{-3}$ as a function of lens galaxy redshift for Classifier-1 (blue) and Classifier-2 (red). The dashed lines indicate the overall TPRs of 0.85 and 0.60 for  Classifier-1 and Classifier-2, respectively. Due to the small sample size, the last redshift bin is chosen to be 0.8--1.1. A histogram of the lens galaxy redshifts of the 92 SuGOHI strong-lens candidates in the test set is also shown (black).}
    \label{fig:roc}
\end{figure*}

\subsection{Classifier-2}

\subsubsection{Training and validation datasets}

As the main focus of this work is finding high-redshift strong lenses, we experimented with a different training set that contains a higher fraction of high-redshift ($z \gtrsim 0.6$) lenses compared to the training set used for Classifier-1. We used the same procedures outlined in Section~\ref{sect:TraningSet} to create mock lenses. The only difference is, at this point we manually adjusted the redshift distribution of the lens galaxies to a relatively uniform distribution from 0.4 to 1.0 (Figure~\ref{fig:TrainingSet_props}) when creating the mocks. Because the number of $z > 0.8$ galaxies in the lens sample is relatively small and each galaxy was only used at most four times, the total number of mock lens systems was 28,500. We therefore augmented the mock lens sample by vertically flipping the cutouts of the 28,500 mock lens systems and considered them as new mock lens systems. 56,960 mock systems were then randomly selected from those 57,000 systems, which we used as the final sample of mock lenses. This new set of mocks has a similar close-to-uniform Einstein radius distribution but clearly contains a higher fraction of higher redshift and fainter lens galaxies, as indicated in Figure~\ref{fig:TrainingSet_props}. Source galaxy properties in these new mocks are not significantly different from those in the mocks for Classifier-1. There is a slightly higher fraction of source galaxies with smaller sizes or bluer $B-V$ and $V-i$ colours, most of which turn out to be at redshifts above 6. For the non-lens examples, we randomly selected another 56,960 objects from the parent sample that do not have counterparts in the known strong lens compilation and the test set for Classifier-1. The randomly shifted $gri$-filter cutouts (60 pixel $\times$ 60 pixel) of the 113,920 mock lenses and non-lens examples are merged into a single dataset, which is again randomly shuffled. 

In addition, two pre-processing steps were introduced. We first took the square root of the absolute value of the dataset. Considering that the lensing features are generally fainter than the lens galaxies, especially in $r$ and $i$ filters, this square-root stretch step improves the contrast between the lens galaxy and lensing features, which has been found to improve the performance of the network (Ca{\~n}ameras et al., in prep.). Afterwards, we normalised the cutouts of every system in the dataset so that the brightest pixel in the individual filter always has a value of 1. Moreover, instead of one network, Classifier-2 is composed of ten networks that are trained with different training sets. This is achieved by implementing the k-fold cross-validation process. More specifically, the single dataset mentioned above was divided into ten chunks of equal size. Each of the ten chunks was used consecutively as the validation set, and the remaining nine chunks were used to train a network. In total, ten networks are obtained, and the average of their output $p_{\rm resnet}$ is used as the final $p_{\rm resnet}$ for every input system.

\subsubsection{Test dataset}

The same 92 lens and 50,000 non-lens examples introduced in Section~\ref{sect:TestSet} were used to construct the test dataset for Classifier-2. The only difference is, their $gri$-filter cutouts also underwent the square-root stretch and normalisation steps. 

\subsubsection{Network tuning}

Similarly, we considered the following three different options of network parameters {\tt learning\_rate}, {\tt learning\_rate\_steps}, and {\tt n\_epochs}: $[0.001, 3, 120]$, $[0.01, 4, 160]$, and $[0.1, 5, 150]$. The set of ten networks that were trained with $[0.1, 5, 150]$ delivered the highest AUROC on the test dataset, and these were chosen as the final networks for Classifier-2. 


\begin{figure*}
    \centering
    \includegraphics[width=0.48\textwidth]{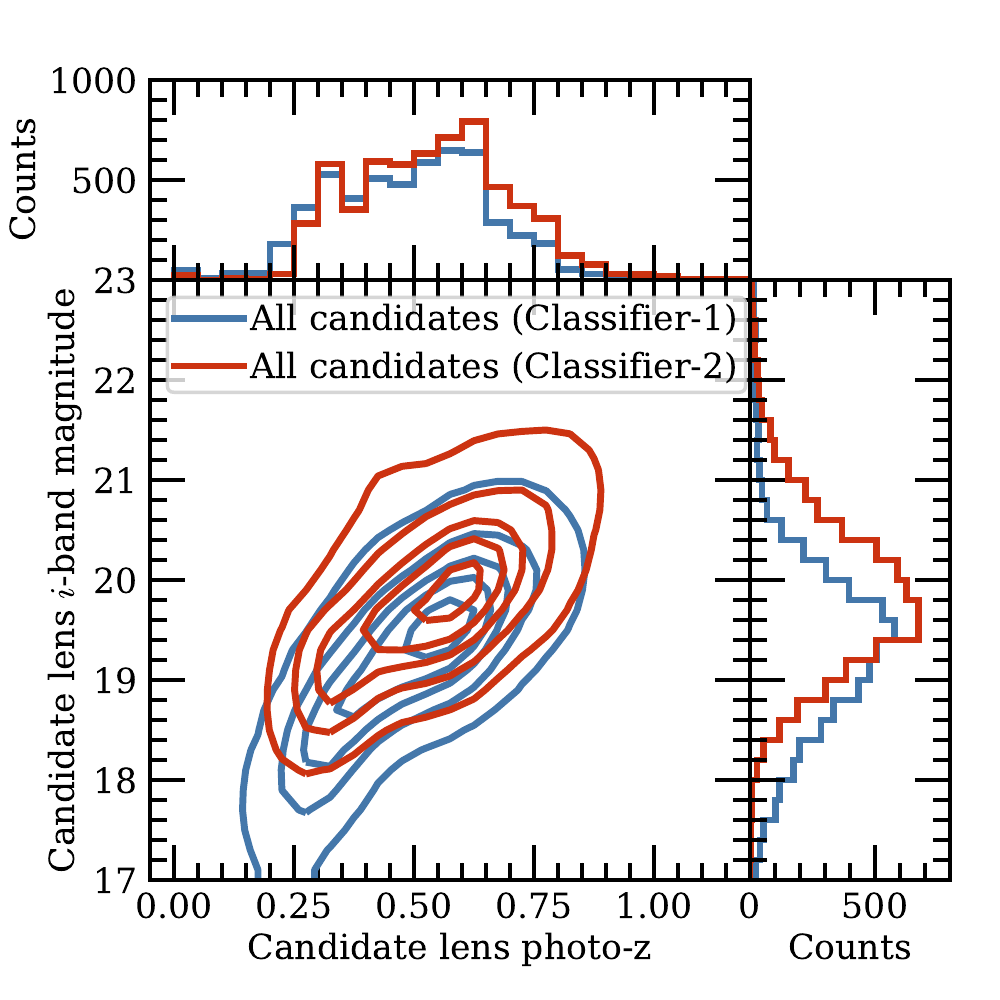}
    \includegraphics[width=0.48\textwidth]{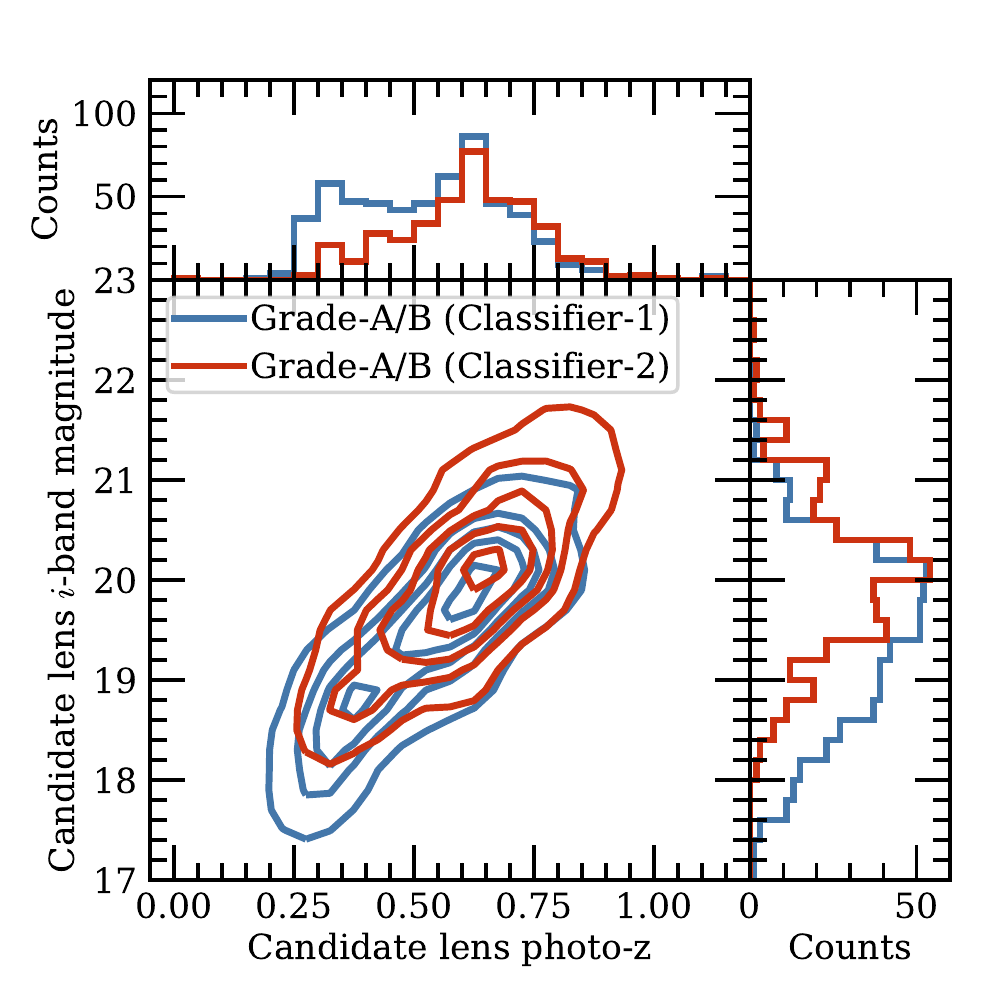}
    \caption{Distributions of photometric redshift and $i-$band magnitude of the lens galaxies in strong-lens candidates found by our two classifiers (left) and the sub-samples that are classified as grade-A or grade-B after visual inspections (right). In both panels, the contours correspond to 10th, 30th, 50th, 70th, and 90th percentiles of the individual dataset.}
    \label{fig:lens_props}
\end{figure*}

\section{Classifier performances}
\label{sect:performance}

Figure~\ref{fig:roc} shows the ROC curves for Classifier-1 (blue) and Classifier-2 (red) based on the test dataset. Classifier-1 has an AUROC of 0.993 and Classifier-2 has an AUROC of 0.985. For reference, the highest AUROC reported in the strong gravitational lens finding challenge was 0.98 \citep{Metcalf19}. \citet{Canameras20} obtained an AUROC of 0.985 and \citet{Huang21} obtained an AUROC of 0.992. Although the AUROC values from different work cannot be directly compared because they are evaluated on different test sets, our AUROC values being in the ballpark of the highest values achieved by recent strong lens classifiers based on neural networks suggests that our two classifiers have been well trained.

For each classifier, we selected a $p_{\rm resnet}$ threshold that delivers an FPR of $10^{-3}$ as the fiducial threshold. Considering the typical strong-lensing rate of $10^{-4}$--$10^{-3}$ \citep[e.g.][]{Browne03, Bolton04, OM10, Treu10}, an FPR of $10^{-3}$ can ensure a reasonable balance between true positives and false positives. In addition, $\approx 6000$ objects in our parent sample (with $\approx$5.36 million objects) are expected to pass the $p_{\rm resnet}$ threshold, which is still manageable in terms of visual inspections. For Classifier-1, the threshold is $p_{\rm resnet} = 0.9731$ and the corresponding TPR is 0.85. For Classifier-2, the threshold is $p_{\rm resnet} = 0.987$ and the corresponding TPR is 0.60. Breaking down into individual redshift bins, we find that the TPRs at an FPR of $10^{-3}$ for Classifier-1 are in agreement with its overall TPR of 0.85 for lens galaxy redshifts from 0.2 to 0.7, beyond which point it drops substantially to TPR $= 0.25$ in the redshift bin of 0.8--1.1 (Figure~\ref{fig:roc}). For Classifier-2, the TPRs for lens galaxy redshifts from 0.2 to 0.4 are lower than its overall TPR of 0.65, presumably because there is no lens galaxy in the training set that is below the redshift of 0.4 for
Classifier-2. The TPR reaches the overall TPR level of 0.65 after the redshift of 0.4 and keeps increasing to almost 0.90 in the redshift bin of 0.7--0.8. In the redshift bin of 0.8--1.1, the TPR for Classifier-2 is 0.50. It becomes clear that even though the overall TPR for Classifier-2 is lower compared to Classifier-1, Classifier-2 is expected to outperform Classifier-1 in discovering strong-lens candidates with lens galaxy redshifts above 0.7. As is shown in the next section, this is further supported by the fact that Classifier-2 has discovered more high-redshift strong-lens candidates from the same parent sample.

\begin{figure*}
    \centering
    \includegraphics[width=0.96\textwidth]{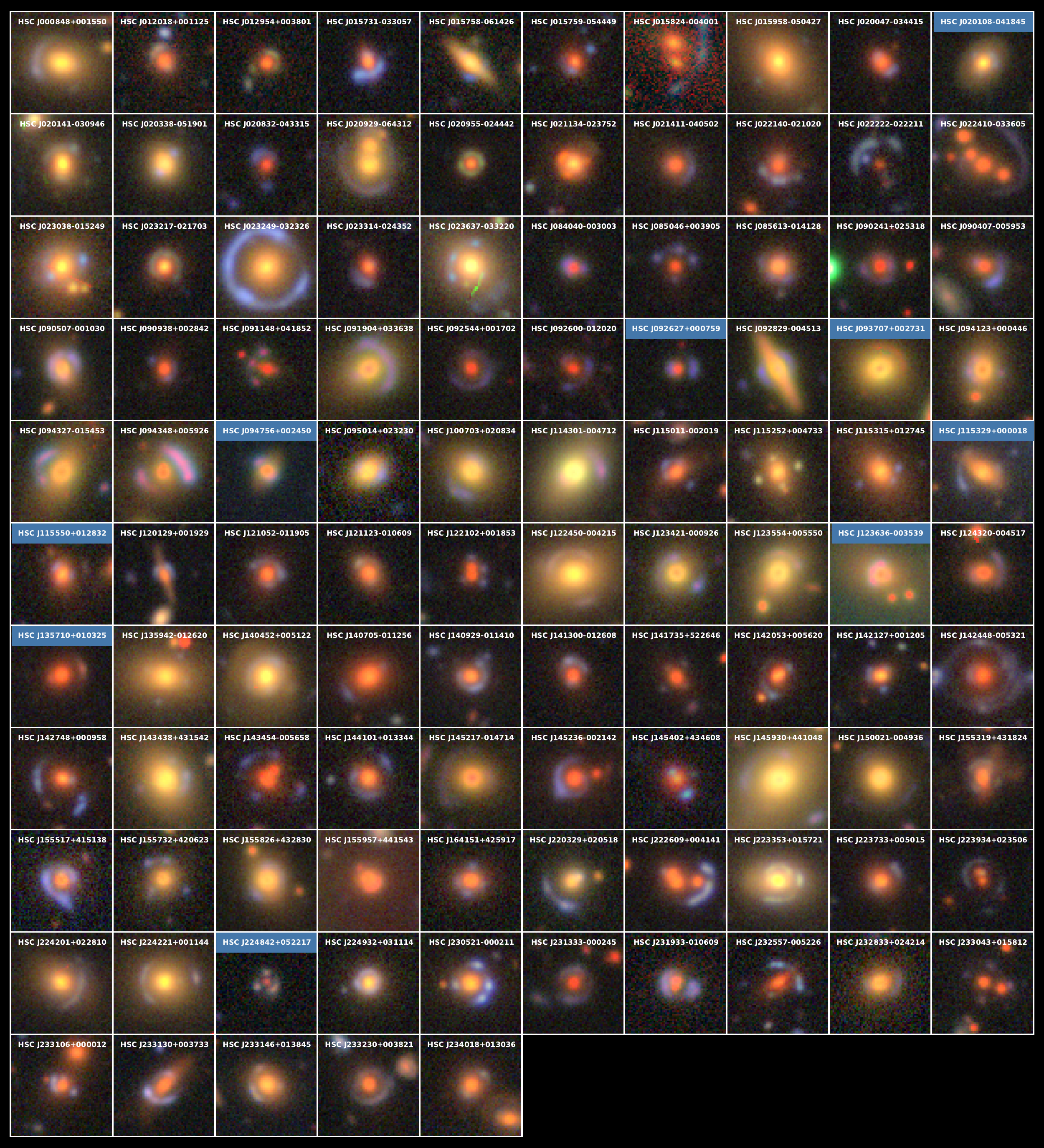}
    \caption{Colour composite images ($10^{\prime \prime} \times 10^{\prime \prime}$) of the 105 grade-A strong-lens candidates discovered by this work. Candidates with a blue background beneath the system name are new discoveries.}
    \label{fig:gradeA}
\end{figure*}

\section{Strong lens candidates in the HSC}
\label{sect:results}

\subsection{Candidates from Classifier-1}

Applying Classifier-1 to our parent sample returned 5,468 unique objects with $p_{\rm resnet} \geq 0.9731$. This fraction, that is 5,468/5,356,628$=$0.00102, is consistent with the FPR of $10^{-3}$ inferred from the test set, which suggests that Classifier-1 is not over-fitted. Those 5,468 objects were considered as strong-lens candidates and passed to visual inspections. The photometric redshift and $i$-band magnitude distributions for the candidate lens galaxies are shown in Figure~\ref{fig:lens_props} (red contours). 

For the visual inspections, author Y. S. performed an initial check of all the 5,468 objects and removed 1,479 obvious non-lenses, which are mostly spiral galaxies, clearly isolated objects, and artefacts. Five authors (Y. S., R. C., S. S., S. H. S., and S. T.) then independently inspected the colour composite cutouts ($10^{\prime \prime} \times 10^{\prime \prime}$, constructed from $gri$ filters) with different scaling schemes and contrasts for the remaining 3,989 objects and assigned an integer score between 0 and 3 to each system following the criteria adopted in \citet{sugohi_I}, \citet{Canameras20}, and \citet{Canameras21}. Specifically, score {\tt 3} corresponds to definite lenses with clear multiple images in configurations that a lens model can easily reproduce. Score {\tt 2} corresponds to probable lenses that have extended and distorted arcs but no clear signs of counter-images and/or would require a lens model to explain the configuration. Score {\tt 1} corresponds to possible lenses with single arcs far away from the central galaxy, and score {\tt 0} corresponds to non-lenses including spirals, ring galaxies, and everything else. The standard deviation of the scores from the five graders was computed for every system. We note that objects with high standard deviations usually show ambiguous arc-like features, which can be interpreted as either lensed background sources or spiral arms of the central galaxies. 531 objects with standard deviations above 0.75 were therefore re-graded by the five graders. 

The visual-inspection scores were averaged over the five graders. 92 systems with average scores $\langle S \rangle \geq {\tt 2.5}$ are considered as grade-A strong-lens candidates and 468 systems with ${\tt 1.5} \leq \langle S \rangle < {\tt 2.5}$ are considered as grade-B strong-lens candidates. Among the 5,468 systems that were inspected, there are 78 grade-A or B SuGOHI galaxy-galaxy strong-lens candidates (again excluding candidates from \citet{sugohi_VI} for the sake of simplicity), and 71 of them have average scores $\langle S \rangle \geq {\tt 1.5}$. The recall of our visual-inspection procedure is therefore estimated to be 91\%. The photometric redshift and $i-$band magnitude distributions for the lens galaxies in the 560 grade-A or B candidates are also shown in Figure~\ref{fig:lens_props}. Among them, 216 (39\%) grade-A or B candidates contain lens galaxies at $z_{\rm d}^{\rm phot} \geq 0.6$ and 22 (4\%) grade-A or B candidates contain lens galaxies at $z_{\rm d}^{\rm phot} \geq 0.8$.   

\subsection{Candidates from Classifier-2}

Applying Classifier-2 to our parent sample returned 6,119 unique objects with $p_{\rm resnet} \geq 0.987$, which is also consistent with the expectation of FPR=$10^{-3}$. Among the 6,119 candidates, 804 were also found by Classifier-1, so their visual-inspection scores were directly set to values from the previous round. Author Y. S. inspected the remaining 5,315 candidates and removed 4,175 candidates that appeared to be non-lenses. The remaining 1,140 candidates were inspected by the same five graders independently. 233 candidates with standard deviations above 0.75 and average score above 1.0 were re-graded. Afterwards, the average visual-inspection scores were computed. In total, Classifier-2 discovers 69 grade-A ($\langle S \rangle \geq {\tt 2.5}$) and 337 grade-B (${\tt 1.5} \leq \langle S \rangle < {\tt 2.5}$) strong-lens candidates. Among the 6,119 systems that were inspected, there are 55 grade-A or B SuGOHI galaxy-galaxy strong-lens candidates, and 51 of them have average scores $\langle S \rangle \geq {\tt 1.5}$. It confirms once again that the recall of our visual-inspection procedure is $\approx$92\%.

Compared to Classifier-1, all 6119 candidates and the 406 grade-A or B candidates found by Classifier-2 tend to contain a higher fraction of higher-redshift or fainter lens galaxies (Figure~\ref{fig:lens_props}). There are 236 (58\%) grade-A or B candidates with lens galaxies at $z_{\rm d}^{\rm phot} \geq 0.6$ and 32 (8\%) grade-A or B candidates with lens galaxies at $z_{\rm d}^{\rm phot} \geq 0.8$. This confirms the finding in the previous Section that Classifier-2 is more effective in discovering strong-lens systems with high-redshift or faint lens galaxies. The reported photometric redshift for one grade-B strong-lens candidate, HSC\,J100400$+$010320, is zero, which is believed to be a catastrophic outlier in the photometric-redshift estimation after checking its image. 

\subsection{The combined sample}

Combining candidates from the two classifiers, we discover in total 105 grade-A and 630 grade-B strong-lens candidates, of which 56 grade-A and 175 grade-B candidates are found by both classifiers. Cross-matching with the known strong lens compilation suggests that 9 grade-A and 268 grade-B candidates are new discoveries. Figure~\ref{fig:gradeA} shows the colour composite images of the 105 grade-A candidates, with the new discoveries indicated by a blue background beneath the system name. Colour composite images of all grade-B candidates are shown in Figure~\ref{fig:gradeB}. Lists of all grade-A and grade-B candidates are presented in Table~\ref{tb:gradeA} and Table~\ref{tb:gradeB}.

There is considerable diversity in the lens and source populations in the discovered grade-A or B strong-lens candidates. The majority of them consist of a single elliptical lens galaxy surrounded by blue, extended lensing-like features, indicating star-forming source galaxies. Nonetheless, some candidates contain disc lens galaxies; for example, HSC\,J015758$-$061426, HSC\,J092829$-$004513, and HSC\,J144228$+$002105. Some candidates show orange or red lensing-like features from source galaxies with overall old stellar populations and/or noticeable dust attenuation; for example, HSC\,J021134$-$023752, HSC\,J093707$+$002731, and HSC\,J155957$+$441543. Some candidates show multiple lensed background sources as being compact; for example, HSC\,J115252$+$004733, HSC\,J122102$+$001853, and HSC\,J224842$+$052217. In addition, there are also some group-scale strong-lens candidates; for example, HSC\,J015824$-$004001, HSC\,J022410$-$033605, and HSC\,J222609$+$004141. 

Nearly half of the discovered grade-A or B strong-lens candidates (331/735) contain lens galaxies with $z_{\rm d}^{\rm phot} \geq 0.6$, of which 4 grade-A and 129 grade-B candidates are new discoveries. 42 candidates contain lens galaxies with $z_{\rm d}^{\rm phot} \geq 0.8$, of which 1 grade-A and 12 grade-B candidates are new discoveries. According to Figure~\ref{fig:lens_props}, the candidate lens galaxies cover a broad magnitude range of 1--2 mag at a fixed redshift, indicating a span of 0.4--0.8 dex in lens galaxy mass. 

\begin{figure}
    \centering
    \includegraphics[width=0.48\textwidth]{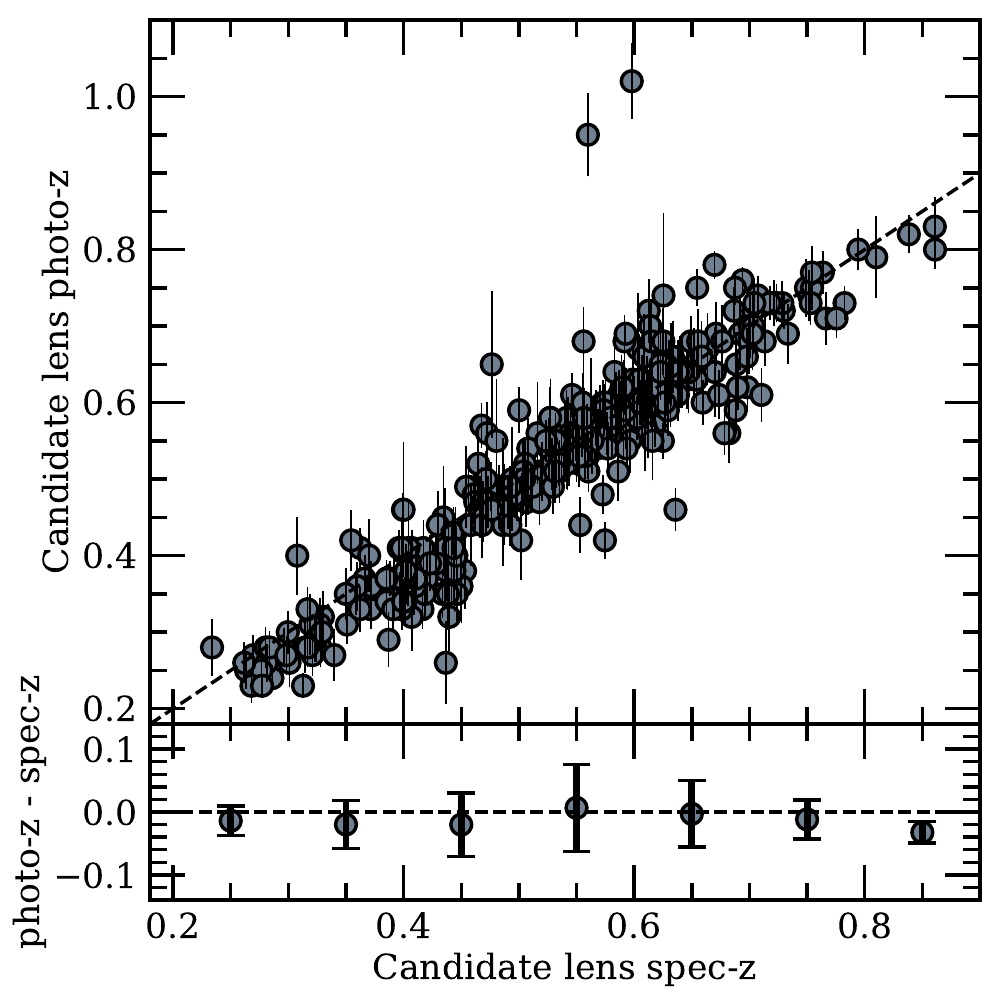}
    \caption{Comparison between photometric redshifts and spectroscopic redshifts for 333 candidate lens galaxies that have measured spectroscopic redshifts (\emph{Top}). The dashed black line is the one-to-one line. The mean and standard deviation of the differences between photometric redshifts and spectroscopic redshifts in seven redshift bins are shown in the bottom panel.}
    \label{fig:redshift_comparison}
\end{figure}

\subsection{Auxiliary spectroscopic data}

We cross-matched our 735 grade-A or B strong-lens candidates with spectroscopic catalogues from SDSS-I \citep{DR7}, SDSS-III \citep{DR12}, SDSS-IV \citep{DR16}, the Master Lens Database$^1$, the SuGOHI project website$^2$, and a sample of spectroscopically-selected strong-lens candidates from \citet{Talbot21} using a matching radius of 1\farcs0, and we obtained spectroscopic redshifts for lens galaxies in 333 candidates and spectroscopic redshifts for source galaxies in 29 candidates. The HSC photometric redshifts for the 333 candidate lens galaxies are in excellent agreement with the corresponding spectroscopic redshifts in general. The differences between the photometric redshifts and spectroscopic redshifts have a mean of $-$0.008 and standard deviation of 0.06 in the redshift range of 0.23--0.86. Divided into seven redshift bins, the mean differences range from $-$0.032 to 0.007 (Figure~\ref{fig:redshift_comparison}), smaller than the average photometric-redshift uncertainty of 0.036 for these 333 galaxies. Photometric redshifts for two candidate lens galaxies, HSC\,J000020$-$002051 (grade-B) and HSC\,J155957$+$441543 (grade-A), are significantly higher than the spectroscopic redshifts (by more than 0.3). For HSC\,J000020$-$002051, the potential lensing features are $\approx 3^{\prime \prime}$ away from the candidate lens galaxy, so the HSC photometry should be reasonably accurate. We think its redshift discrepancy is likely due to a catastrophic failure in the photometric-redshift estimation, which is supported by the fact that the photometric redshift for the same galaxy in DESI Legacy Imaging Surveys Data Release 9 is $0.59 \pm 0.03$ \citep{Dey19}, in agreement with the spectroscopic redshift of 0.560. For HSC\,J155957$+$441543, we think the photometric redshift is biased high due to the contamination from the candidate source galaxy, which is red in colour and is comparably as bright as the candidate lens galaxy in all five HSC filters. The photometric redshift for the same galaxy from the DESI Legacy Imaging Surveys is also over-estimated as $0.72 \pm 0.10$. Nevertheless, the overall agreement suggests that the photometric-redshift estimation for candidate lens galaxies in our sample is barely affected by the presence of surrounding potential lensing features. This is understandable as our visual inspection process preferentially picks out candidates that exhibit clear separations between the central galaxies and potential lensing features. Moreover, CModel photometry, instead of aperture photometry, is used for the photometric-redshift estimation \citep{Tanaka2018}, in which substantial deblending from surrounding features is already involved. It also indicates that the photometric redshifts for the remaining candidate lens galaxies are likely reliable. 
\begin{figure*}
    \centering
    \includegraphics[width=0.96\textwidth]{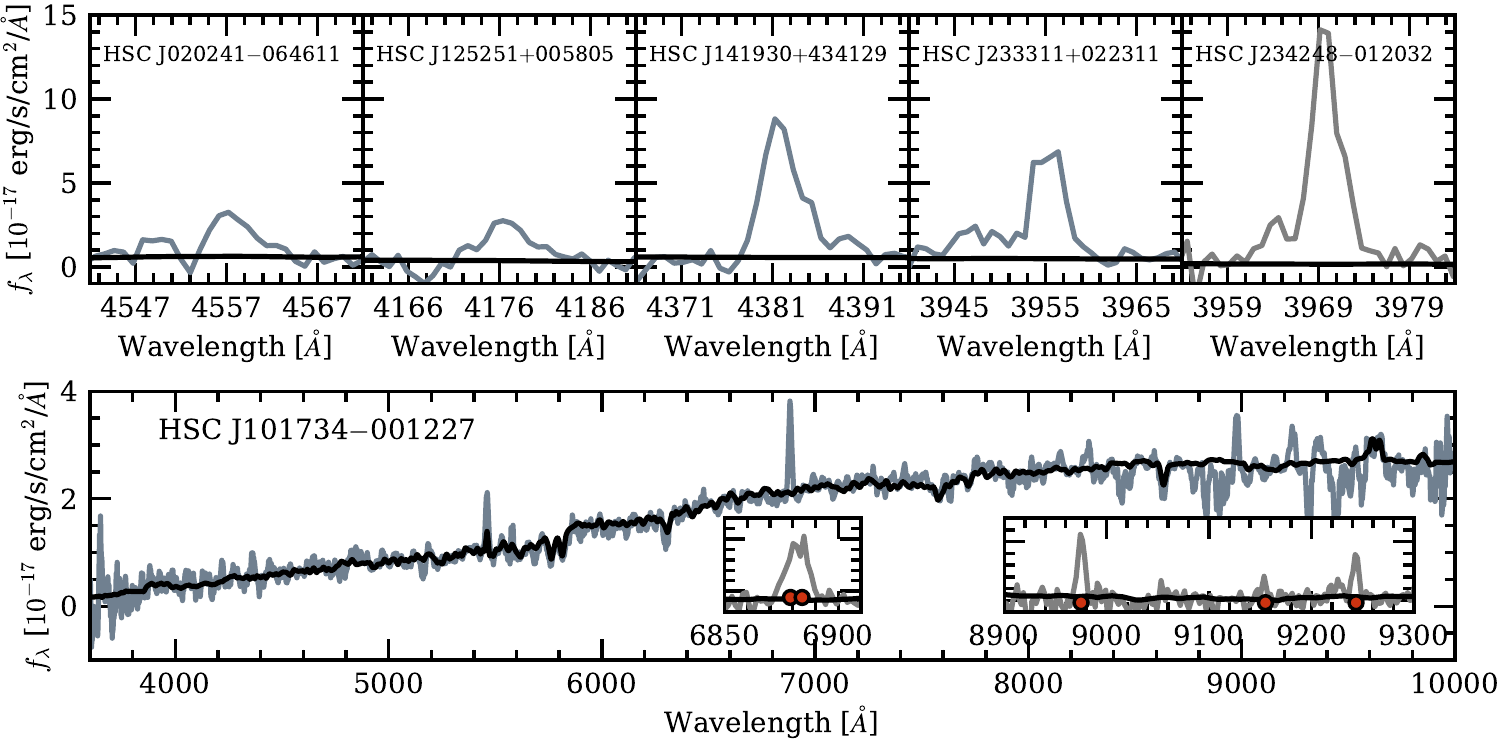}
    \caption{SDSS spectra of the six strong-lens candidates with evidence of higher-redshift emission lines. In each panel, the grey line represents the observed spectrum and the black line represents the SDSS-provided best-fit model spectrum (only for the foreground lens galaxy). The top row shows 30\AA \ windows centred on the detected emission line for HSC\,J020241$-$064611, HSC\,J125251$+$005805, HSC\,J141930+434129, HSC\,J233311$+$022311, and HSC\,J234248$-$012032, and the bottom row shows the full optical spectrum for HSC\,J101734-001227. Several emission lines not associated with the redshift of the foreground galaxy (i.e. $z=0.4647$) are shown in the zoomed-in images in the insets of the bottom panel. They are found to be coincident with the locations of [O\textsc{ii}] doublet, H$\beta$, [O\textsc{iii}]~4960, and [O\textsc{iii}]~5008 at $z=0.8457$.}
    \label{fig:candidates_spectra}
\end{figure*}

\section{Notes on individual systems}
\label{sect:notes}

We carried out visual inspections of the publicly available spectra of the 333 candidates identified in the previous section and found six cases where prominent emission lines not consistent with the redshift of the candidate lens galaxies are detected, suggesting superpositions of two objects along the same line of sight. We discuss those cases one by one in this section. We note, however, that this list is by no means complete, and interested readers are encouraged to conduct their own analyses.

\subsection{HSC\,J020241$-$064611}
This is a grade-B candidate according to our visual inspection. Two blue, arc-like features are found on the north and south sides of an orange, elliptical galaxy with a separation of $\approx$1\farcs8 (Figure~\ref{fig:gradeB}). A fibre-fed (2$^{\prime \prime}$ in diameter) spectrum from SDSS-III is available, which shows a high S/N emission line at 4557.2\AA\ on top of a $z=0.5020$ early-type galaxy spectrum (Figure~\ref{fig:candidates_spectra}). This line is obviously not coincident with any typical emission line at $z=0.5020$. \citet{Shu16} interpreted this line as Ly$\alpha$ emission from a Ly$\alpha$ emitter (LAE) at $z=2.7477$, and considered this system as a galaxy-LAE strong-lens candidate. This system was also classified, based on HSC data, as a grade-B candidate by \citet{sugohi_I}, who resolved the two arc-like features after subtracting the foreground galaxy light. Combining imaging and spectroscopic evidence, we speculate that the two arc-like features are lensed images of a $z=2.7477$ LAE. The SDSS-measured central velocity dispersion for the foreground galaxy is $156 \pm 25$ km\,s$^{-1}$ \footnote{Starting from Data Release 9, SDSS provides two types of velocity dispersion. One is {\tt VDISP} determined by fitting the observed spectrum with a linear combination of 24 eigenspectra. The other can be inferred from {\tt VDISP\_LNL}, which is the velocity-dispersion likelihood function computed by fitting with a linear combination of five eigenspectra while marginalising over redshift uncertainties. As discussed in \citet{Shu12} and \citet{Bolton12}, velocity dispersions inferred from {\tt VDISP\_LNL} are more robust for SDSS-III galaxies, the spectra of which often have relatively low S/N. We therefore adopt the velocity dispersion inferred from {\tt VDISP\_LNL} in this work.}, which corresponds to an Einstein radius of $\approx 0\farcs48 \pm 0\farcs15$ for a source at $z=2.7477$ and a lens at $z=0.5020$ with an isothermal total-mass profile. The estimated Einstein radius is $\approx 2.8\sigma$ lower than what is suggested from the image separation. 

\subsection{HSC\,J101734-001227}
This is a grade-B candidate according to our visual inspection and was also classified as grade-B by C21. A red, elongated arc is located $\approx 1\farcs6$ west of an orange, elliptical galaxy, and there seems to be some hint of a counter image very close to the elliptical galaxy (Figure~\ref{fig:gradeB}). A fibre-fed (2$^{\prime \prime}$ in diameter) spectrum from SDSS-III is available. The SDSS best-fit model suggests a redshift of 0.8457, which is primarily driven by several strong emission lines being coincident with [O\textsc{ii}] doublet, H$\beta$, [O\textsc{iii}]~4960, and [O\textsc{iii}]~5008 at $z=0.8457$. Nevertheless, it is noticed that some emission and absorption features in the spectrum cannot be explained by the best-fit model. Interestingly, we find that the second-best fit using galaxy templates at $z=0.4647$ provided by SDSS can well reproduce those emission and absorption features (Figure~\ref{fig:candidates_spectra}). It hence becomes clear that this particular line of sight contains two galaxies, one at $z=0.4647$ and the other at $z=0.8457$. Unfortunately we cannot estimate the Einstein radius because the SDSS-reported velocity dispersion is $850$ km\,s$^{-1}$, indicating a failure in the measurement. Combining imaging and spectroscopic evidence, we speculate that the potential counter image and/or the elongated arc on the west are responsible for the detected [O\textsc{ii}] doublet, H$\beta$, [O\textsc{iii}]~4960, and [O\textsc{iii}]~5008 at $z=0.8457$. 

\subsection{HSC\,J125251$+$005805}
This is a grade-B candidate according to our visual inspection. A blue, elongated arc and a similarly blue blob are found on the northeast and southwest sides of an orange, elliptical galaxy with a separation of $\approx 1\farcs9$. A fibre-fed (2$^{\prime \prime}$ in diameter) spectrum from SDSS-III is available, which shows a high S/N emission line at 4176.4\AA\ on top of a $z=0.5399$ early-type galaxy spectrum (Figure~\ref{fig:candidates_spectra}). This line is obviously not coincident with any typical emission line at $z=0.5399$. \citet{Shu16} interpreted this line as Ly$\alpha$ emission from an LAE at $z=2.4345$, and considered this system as a galaxy-LAE strong-lens candidate. This system was also classified, based on HSC data, as a grade-B candidate by \citet{sugohi_II}. The SDSS-measured central velocity dispersion for the foreground galaxy is $203 \pm 40$ km\,s$^{-1}$, which corresponds to an Einstein radius of $\approx 0\farcs8 \pm 0\farcs3$ for a source at $z=2.4345$ and a lens at $z=0.5399$ with an isothermal total-mass profile. The estimated Einstein radius is in good agreement with the observed image separation. Combining imaging and spectroscopic evidence, we think that the blue arc and blob are indeed lensed images (in a cusp configuration) of a $z=2.4345$ LAE. 

\subsection{HSC\,J141930+434129}
This is a grade-B candidate according to our visual inspection. A blue, elongated arc is located $\approx 1\farcs5$ southwest of an orange, elliptical galaxy, but there is no decisive sign for any counter image in the HSC data (Figure~\ref{fig:gradeB}). A fibre-fed (2$^{\prime \prime}$ in diameter) spectrum from SDSS-IV is available, which shows a high S/N emission line at 4381.3\AA\ on top of a $z=0.5447$ early-type galaxy spectrum (Figure~\ref{fig:candidates_spectra}). We verified that the detected line is present in the 1D spectra from three individual sub-exposures. This line is obviously not coincident with any typical emission line at $z=0.5447$. It is also unlikely to be a low-redshift [O\textsc{ii}] doublet, because no other strong emission is detected at wavelength positions that would correspond to H$\beta$, [O\textsc{iii}], and H$\alpha$. We hence interpret this line as Ly$\alpha$ emission at $z=2.6030$. The SDSS-measured central velocity dispersion for the foreground galaxy is $200 \pm 40$ km\,s$^{-1}$, which corresponds to an Einstein radius of $\approx 0\farcs8 \pm 0\farcs3$ for a source at $z=2.6030$ and a lens at $z=0.5447$ with an isothermal total-mass profile. Combining imaging and spectroscopic evidence, we speculate that the detected Ly$\alpha$ emission is primarily from the blue arc on the southwest (due to scattering). If there is indeed a faint counter image close to the foreground galaxy, which is consistent with the Einstein radius estimation, it would also contribute to the detected Ly$\alpha$ emission. 

\subsection{HSC\,J233311$+$022311.}
This is a grade-B candidate according to our visual inspection. Two tangentially elongated blue blobs are located $\approx 1\farcs3$ southeast of an orange, elliptical galaxy, and there is no sign for any counter image in the HSC data (Figure~\ref{fig:gradeB}). A fibre-fed (2$^{\prime \prime}$ in diameter) spectrum from SDSS-III is available, which shows a strong emission line at 3955.5\AA\ on top of a $z=0.4716$ early-type galaxy spectrum (Figure~\ref{fig:candidates_spectra}). This line is obviously not coincident with any typical emission line at $z=0.4716$. \citet{Shu16} interpreted this line as Ly$\alpha$ emission from an LAE at $z=2.2529$, and considered this system as a galaxy-LAE strong-lens candidate. This system was also classified, based on HSC data, as a grade-B candidate by \citet{sugohi_II}. The SDSS-measured central velocity dispersion for the foreground galaxy is $272 \pm 55$ km\,s$^{-1}$, which corresponds to an Einstein radius of $\approx 1\farcs4 \pm 0\farcs6$ for a source at $z=2.2529$ and a lens at $z=0.4716$ with an isothermal total-mass profile. Combining imaging and spectroscopic evidence, we speculate that the detected Ly$\alpha$ emission is primarily from the two blue blobs on the southeast. If there is indeed a faint counter image close to the foreground galaxy, which is broadly consistent with the Einstein radius estimation, it would also contribute to the detected Ly$\alpha$ emission. 

\subsection{HSC\,J234248$-$012032.}
This is a grade-B candidate according to our visual inspection. A blue, elongated arc and a similarly blue blob are found on the northwest and southeast sides of an orange, elliptical galaxy with a separation of $\approx 2\farcs1$. A fibre-fed (2$^{\prime \prime}$ in diameter) spectrum from SDSS-III is available, which shows a high S/N emission line at 3970.1\AA\ on top of a $z=0.5270$ early-type galaxy spectrum (Figure~\ref{fig:candidates_spectra}). This line is obviously not coincident with any typical emission line at $z=0.5270$. \citet{Shu16} interpreted this line as Ly$\alpha$ emission from an LAE at $z=2.2649$, and considered this system as a galaxy-LAE strong-lens candidate. The SDSS-measured central velocity dispersion for the foreground galaxy is $271 \pm 44$ km\,s$^{-1}$, which corresponds to an Einstein radius of $\approx 1\farcs4 \pm 0\farcs4$ for a source at $z=2.2649$ and a lens at $z=0.5270$ with an isothermal total-mass profile. The estimated Einstein radius is in good agreement with the observed image separation. Combining imaging and spectroscopic evidence, we think that the blue arc and blob are indeed lensed images (in a cusp configuration) of a $z=2.2649$ LAE. 

\section{Discussions}
\label{sect:discussion}

As already demonstrated in Section~\ref{sect:performance} and Section~\ref{sect:results}, Classifier-2 is more effective than Classifier-1 in the discovery of strong-lens systems with high-redshift or faint lens galaxies, which, essentially, is a result of differences in the training set and pre-processing steps. 60\% and 28\% of the mock lenses used for Classifier-2 are at redshifts above 0.6 and fainter than $i=20.5$ mag, respectively, while these two fractions are only 24\% and 4\% for Classifier-1. In addition, the square-root stretch implemented only in Classifier-2 helps to better reveal lensing features in high-redshift lenses, which, by construction, require higher-redshift sources that appear fainter on average than sources in lower-redshift lenses. Interestingly, we find that including the two pre-processing steps (square-root stretch and normalisation) in Classifier-1 or removing them from Classifier-2 leads to worse performance in terms of AUROC. These findings highlight that the outcome of supervised machine learning techniques depends strongly on the training set and pre-processing procedures need to be chosen in accordance with the training set. We tested training classifiers with $griz$-filter (instead of $gri$) cutouts, but the performance was not as good as the two presented classifiers. More thorough discussions on the impact of the training set will be presented in Ca{\~n}ameras et al. (in prep.) and More et al. (in prep.). 

According to the 0.85 TPR for Classifier-1 and $\approx$92\% visual-inspection recall, the 560 grade-A or B strong-lens candidates discovered by Classifier-1 suggest that, in our parent sample, there would be 716 strong lenses in total with properties similar to the 92 SuGOHI strong-lens candidates in our test set. Likewise, the 406 grade-A or B strong-lens candidates discovered by Classifier-2 with a TPR of 0.60 suggest a total number of 736 strong lenses. These two predictions agree well with each other, and they are also consistent with the 735 grade-A or B strong-lens candidates discovered by the two classifiers combined. From another perspective, 84 of the 92 SuGOHI candidates in our test set are recovered by the two classifiers combined, suggesting an overall recall of 91\%. We therefore expect that $\gtrsim$90\% of all strong-lens candidates that are in our parent sample and have properties similar to the 92 SuGOHI strong-lens candidates have already been included in our lists of grade-A or B strong-lens candidates. 

\begin{figure}
    \centering
    \includegraphics[width=0.48\textwidth]{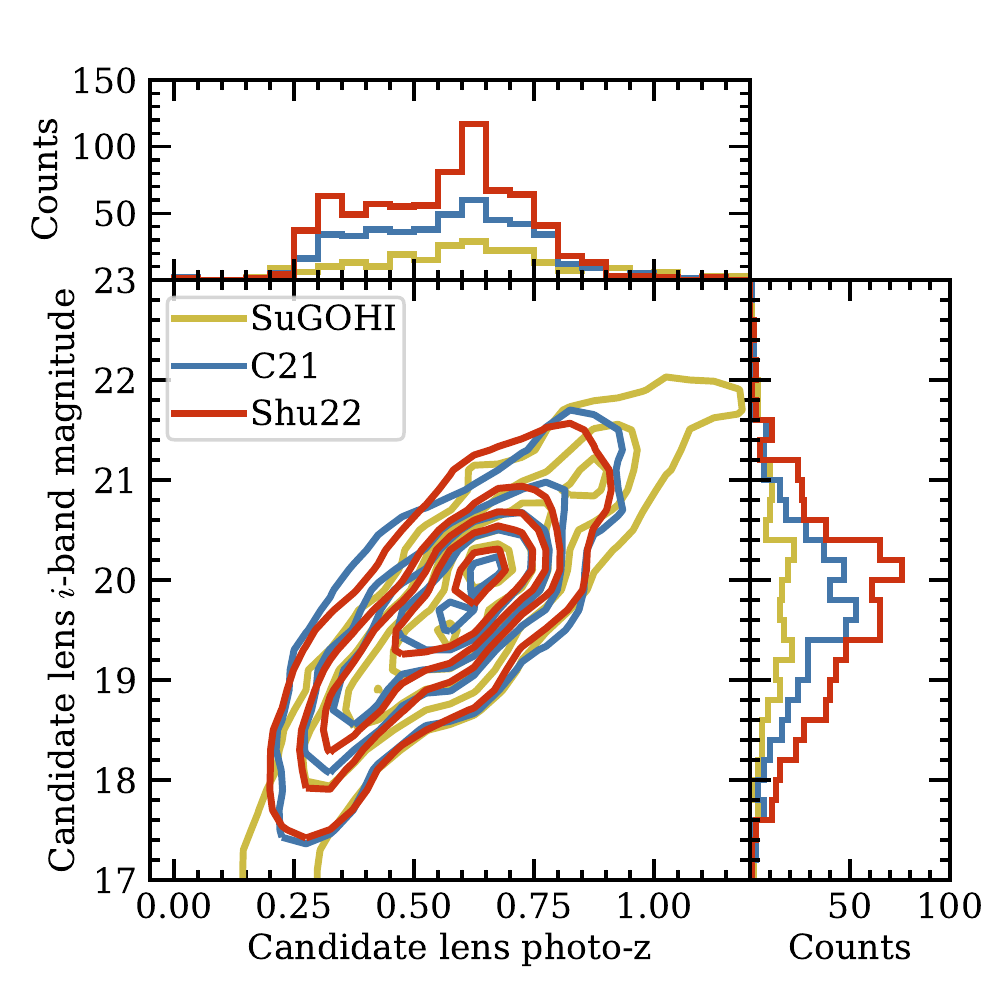}
    \caption{Distributions of photometric redshift and $i-$band magnitude of the lens galaxies in galaxy-scale strong-lens candidates from the SuGOHI project (yellow), C21 (blue), and this work (red). The contours correspond to 10th, 30th, 50th, 70th, and 90th percentiles of the individual samples. 
    To make a fair comparison, we use {\tt photoz\_best} from the {\tt pdr2\_wide.photoz\_mizuki} table for SuGOHI lens galaxies instead of the photometric redshifts provided by the SuGOHI project website$^2$.}
    \label{fig:comparison}
\end{figure}

\citet{Collett15} made a prediction on the population of detectable galaxy-galaxy strong lenses in several imaging surveys. Although HSC was not considered there, we can use results for the LSST, relevant properties of which (including pixel scale, seeing distribution, and sky-brightness distribution) are similar to HSC, as an approximation. In particular, \citet{Collett15} forecasted that LSST can detect, over an area of 20,000 deg$^2$, 17,000 galaxy-galaxy strong lenses from the best single-epoch imaging and 39,000 galaxy-galaxy strong lenses from the final full stack of the survey. The nominal depths of LSST single-epoch and full-stack imaging are \{25.0, 24.7, 24.0\} and \{27.4, 27.5, 26.8\} in \{$g$, $r$, $i$\} filters \citep{Ivezic19}, which nicely bracket the depths of HSC PDR2. It hence suggests that the total number of detectable galaxy-galaxy strong lenses in HSC PDR2 is between 800 and 1900. In terms of high-redshift strong lenses, the forecast is that there will be between 180 and 190 $z_{\rm d} > 0.8$ strong lenses. We note that the actual number of detectable strong lenses is very sensitive to the adopted S/N threshold. \citet{Collett15} considered a lens system to be detectable if the total S/N, SN$_{\rm TOT}$, of the lensing features is 20 or higher in at least one band (along with three other conditions). If requiring SN$_{\rm TOT} > 30$, the forecasts for the total number of strong lenses and $z_{\rm d} > 0.8$ strong lenses in HSC PDR2 drop to 300--1200 and 80--110. On the other hand, \citet{Collett15} pointed out that their LSST forecasts are likely underestimated due to poorly constrained redshift and size distributions of source galaxies used in their simulation, especially on the faint end. The uncertainties were estimated to be at the level of $\sim$10\%. It is unclear what fraction of the detectable strong lenses simulated in \citet{Collett15} can pass our selection criteria in Section~\ref{sect:data}. Nevertheless, we believe that the vast majority of our grade-A or B strong-lens candidates have SN$_{\rm TOT}$ substantially higher than 20 according to Figure~\ref{fig:gradeA} and Figure~\ref{fig:gradeB}, and our single set of 735 grade-A or B strong-lens candidates (including 42 at $z_{\rm d}^{\rm phot} > 0.8$) represents $\gtrsim 50\%$ of all detectable strong lenses in HSC PDR2.

Prior to this work, there were two other projects that searched systematically for strong lenses in the HSC data. One of them is the SuGOHI project and the other is a project also done by us, that is C21. The SuGOHI project makes use of several different methods for lens search including automated algorithms \citep[e.g.][]{sugohi_I, sugohi_IV} and crowdsourcing \citep[e.g.][]{sugohi_VI}. C21 makes use of a resnet, similar to this work. Time-wise, the resnets used in C21 and this work are many orders of magnitude faster than the automated algorithms and crowdsourcing used in the SuGOHI project. Classifications of the 5.3 million objects in this work took $\approx 100$ minutes, or $\approx$50,000 objects per minute. The classification speed of the methods used in the SuGOHI project is on the order of $\sim 10$s per object (K. Wong, private communication).

A more fundamental distinction between the three projects is on the parent sample. The parent sample of this work contains galaxies (or more precisely speaking, extended objects) in the Wide layer of HSC PDR2 that satisfy certain magnitude and colour cuts defined in Section~\ref{sect:data} (along with some quality flags). The parent sample in C21 is 62.5 million galaxies in the Wide layer of HSC PDR2 with an $i-$band Kron radius larger than 0\farcs8. The parent samples in the SuGOHI project are more heterogeneous and selected not only from the Wide layer but also the HSC Deep and UltraDeep fields. In particular, the parent samples in \citet{sugohi_I}, \citet{sugohi_II}, and \citet{sugohi_IV} are $\approx 500,000$ luminous red galaxies selected for spectroscopic observations in SDSS-III. The parent sample in \citet{sugohi_VI} is $\approx 300,000$ galaxies with photometric redshifts between 0.2 and 1.2 and stellar mass above $10^{11.2} M_{\odot}$. In our parent sample, 3,493,859 (65.2\%) objects have $i-$band Kron radius smaller than 0\farcs8 and 4,957,066 (92.5\%) objects do not satisfy either of the two requirements in the SuGOHI project. As a result, approximately 3.4 million objects in our parent sample had not been classified by either the SuGOHI project or C21. In terms of high-redshift galaxies, our parent sample is much more complete than those in the other two projects. 80--90\% of HSC PDR2 galaxies at redshifts above 0.8 are expected to be included by the colour-colour cuts in this work. In our parent sample, 1,402,958 objects have photometric redshifts above 0.8, of which only 524,078 (37.4\%) have $i-$band Kron radius larger than 0\farcs8. It suggests that the parent sample in C21 only included approximately one third of all HSC PDR2 galaxies at redshifts above 0.8. The total size of the parent samples in the SuGOHI project is only $\approx$800,000, and redshifts for the vast majority are below 0.8.

The SuGOHI project has discovered 497 grade-A or B strong-lens candidates, of which 248 are classified as galaxy-scale systems. For the following comparisons, galaxy-scale candidates from \citet{sugohi_VI} are also included in this SuGOHI galaxy-scale sample, although some of them are actually cluster- or group-scale systems as pointed out in Section~\ref{sect:data}. C21 has discovered 467 grade-A or B strong-lens candidates, almost all of which are galaxy-scale systems. Similarly, almost all of the 735 grade-A or B strong-lens candidates discovered by this work are galaxy-scale systems. There are 132 candidates in common between this work and the SuGOHI project, and 302 candidates in common between this work and C21. Combining these three sample yields 1,002 unique galaxy-scale strong-lens candidates, and 395 of the 735 (54\%) grade-A or B strong-lens candidates in this work had not been discovered by the other two projects. Candidates in C21 and this work cover similar ranges in lens galaxy photometric redshift and $i-$band magnitude, while the SuGOHI galaxy-scale sample contains a higher fraction of candidates with lens photometric redshifts above $\approx$0.9 (Figure~\ref{fig:comparison}). In terms of numbers, 25 candidates in the SuGOHI galaxy-scale sample, 13 candidates in C21, and 11 candidates in our sample have lens photometric redshifts above $\approx$0.9. Nevertheless, we find that 13 of the 25 $z_{\rm d}^{\rm phot} > 0.9$ SuGOHI galaxy-scale candidates and 8 of the 13 $z_{\rm d}^{\rm phot} > 0.9$ C21 candidates do not fulfil our colour selection criteria defined in Section~\ref{sect:data} (criteria 13--14) and are not included in our parent sample in the first place. Those candidates generally have bluer $g-i$ colours as a result of the contamination from the blue lensing features, especially in the $g$ band. However, their photometric-redshift estimations appear not to be significantly affected by this type of contamination (see also Figure~\ref{fig:redshift_comparison}), likely because the photometric-redshift estimation is based on multiple colours and is therefore less sensitive to any bias in one particular band.

To further improve completeness with regard to discovering high-redshift strong lenses, a few options can be explored. The first is to improve the completeness in the parent sample. Although the colour-colour cuts used in this work are found to be already 80--90\% complete in selecting high-redshift strong lenses, some known high-redshift strong-lens candidates are excluded due to contaminated photometry. On the other hand, the provided photometric redshifts do not seem to be strongly biased by lensing features in general. Combining the colour-colour criteria and a photometric-redshift selection should in principle result in a more complete parent sample. Moreover, the classifier may be further optimised. In this work, the classifiers were tuned to deliver high overall TPRs for strong-lens candidates covering a wide redshift range from 0.2 to 1.1, and it has been shown that the TPRs can vary substantially in different redshift sub-ranges. One can consider optimising the classifier based on the performance on the redshift range of interest only.

\section{Conclusions}
\label{sect:conclusion}

In this work, we carried out a search for strong-lens systems consisting of high-redshift lens galaxies in the Wide layer data from HSC-SSP PDR2 with a sky coverage of $\approx$960 deg$^2$. We first applied several colour and magnitude cuts to reduce the sample size in HSC PDR2 from $\approx 80$ million galaxies to $\approx 5.4$ million galaxies. To further efficiently classify those galaxies, that is our parent sample, we constructed two strong-lens classifiers based on a deep residual network pre-built in the {\tt CMU DeepLens} package. The two classifiers, Classifier-1 and Classifier-2, differ mainly in the training set and pre-processing procedures. After training, the two classifiers achieved AUROC values of 0.993 and 0.985 on a test dataset comprising real strong lenses and non-lenses. Applying each of the two classifiers to the $gri$-filter cutouts (60 pixel $\times$ 60 pixel, 1 pixel = 0\farcs17) of the parent sample returned network scores $p_{\rm resnet}$ for individual galaxies in $\approx 100$ minutes. Adopting $p_{\rm resnet}$ thresholds that correspond to an FPR of $10^{-3}$ based on the test set, Classifier-1 and Classifier-2 produced 5,468 and 6,119 unique strong-lens candidates, respectively. Five authors independently graded those strong-lens candidates based on visual inspections of the cutouts. According to the average visual-inspection scores, 560 candidates identified by Classifier-1 and 406 candidates identified by Classifier-2 are considered as grade-A or B (i.e. definite or probable) strong-lens candidates. 

By combining the two samples, we discover in total 105 grade-A and 630 grade-B strong-lens candidates, which is the single largest set of galaxy-scale strong-lens candidates discovered with HSC data to date. Among them, nine grade-A and 268 grade-B candidates are new discoveries. This list of 735 candidates is expected to include $\gtrsim 90\%$ of all strong-lens candidates that are in our parent sample and have properties similar to the test set. The candidate lens galaxies span a (photometric) redshift range from 0.2 to 1.0. Nearly half of the discovered candidates (331/735) contain lens galaxies with $z_{\rm d}^{\rm phot} \geq 0.6$, and 42 candidates contain lens galaxies with $z_{\rm d}^{\rm phot} \geq 0.8$. Despite having a lower overall TPR, Classifier-2 discovers a significantly higher fraction of high-redshift ($z_{\rm d}^{\rm phot} \geq 0.6$) lens galaxies compared to Classifier-1, which we attribute to differences in the training set and pre-processing procedures. 

We obtained spectroscopic redshifts for lens galaxies in 333 candidates and spectroscopic redshifts for source galaxies in 29 candidates by cross-matching our candidates with spectroscopic catalogues in the literature. We found an excellent agreement between the HSC-reported photometric redshifts and the corresponding spectroscopic redshifts for the 333 candidate lens galaxies, indicating that the photometric redshifts for the remaining candidate lens galaxies are likely reliable. In addition, we noticed high S/N emission lines in publicly-available spectra of six candidates that are presumably from redshifts higher than those of the foreground galaxies. It is worth carrying out follow-up observations to determine the nature of the detected emission lines and lensing status of the six systems.  

We will continue applying our classifiers to future HSC data releases to discover more strong-lens systems. Meanwhile, we will obtain follow-up spectroscopy to confirm the best-quality high-redshift strong-lens candidates from this search and turn them into a powerful probe for galaxy evolution at $z \gtrsim 0.8$. Our discoveries will also serve as a valuable target list for ongoing and scheduled spectroscopic surveys such as the Dark Energy Spectroscopic Instrument \citep{DESI}, the Subaru Prime Focus Spectrograph project \citep{PFS}, and the Maunakea Spectroscopic Explorer \citep{MSE}. As demonstrated by this work, resnet-based algorithms are a promising approach for efficiently and effectively uncovering the $\sim 10^{5}$ strong-lens systems expected in forthcoming wide-field imaging surveys such as LSST, Euclid, and CSS-OS. All kinds of scientific applications enabled by strong lensing are expected to benefit from a larger and more complete population of strong-lens systems.

\begin{acknowledgements}
The authors thank Drs. Thomas Collett and Kenneth Wong for helpful discussions, and the anonymous referee for constructive comments that improved the presentation of this work. Y. S. acknowledges support from the Max Planck Society and the Alexander von Humboldt Foundation in the framework of the Max Planck-Humboldt Research Award endowed by the Federal Ministry of Education and Research. 
S.~H.~S.~thanks the Max Planck Society for support through the Max
Planck Research Group.  
This project has received funding from the European Research Council (ERC)
under the European Union’s Horizon 2020 research and innovation
programme (LENSNOVA: grant agreement No 771776).
This research is supported in part by the Excellence Cluster ORIGINS which is funded by the Deutsche Forschungsgemeinschaft (DFG, German Research Foundation) under Germany's Excellence Strategy -- EXC-2094 -- 390783311.
A. T. J. is supported by P2MI ITB 2021. 
The Hyper Suprime-Cam (HSC) collaboration includes the astronomical communities of Japan and Taiwan, and Princeton University. The HSC instrumentation and software were developed by the National Astronomical Observatory of Japan (NAOJ), the Kavli Institute for the Physics and Mathematics of the Universe (Kavli IPMU), the University of Tokyo, the High Energy Accelerator Research Organization (KEK), the Academia Sinica Institute for Astronomy and Astrophysics in Taiwan (ASIAA), and Princeton University. Funding was contributed by the FIRST program from the Japanese Cabinet Office, the Ministry of Education, Culture, Sports, Science and Technology (MEXT), the Japan Society for the Promotion of Science (JSPS), Japan Science and Technology Agency (JST), the Toray Science Foundation, NAOJ, Kavli IPMU, KEK, ASIAA, and Princeton University. This paper makes use of software developed for the Large Synoptic Survey Telescope. We thank the LSST Project for making their code available as free software at  http://dm.lsst.org. This paper is based [in part] on data collected at the Subaru Telescope and retrieved from the HSC data archive system, which is operated by the Subaru Telescope and Astronomy Data Center (ADC) at National Astronomical Observatory of Japan. Data analysis was in part carried out with the cooperation of Center for Computational Astrophysics (CfCA), National Astronomical Observatory of Japan. The Subaru Telescope is honored and grateful for the opportunity of observing the Universe from Maunakea, which has the cultural, historical and natural significance in Hawaii.

\end{acknowledgements}


\begin{appendix} 
\section{Visual-inspection score comparisons with C21}

Among all the network candidates from C21 and this work, 956 systems are in common and have been assigned visual-inspection scores twice by the same five graders. In this appendix, we discuss the variations in the visual-inspection scores for the same systems from round to round, which provides an idea on the robustness of our visual-inspection scores. We note that the visual-inspection processes between this work and C21 are slightly different. In C21, three images with different stretching and normalisation schemes for the same systems were provided to the graders, while four more images with different stretching and normalisation schemes for the same systems were provided in this work. 

\begin{figure*}[h]
    \centering
    \includegraphics[width=0.9\textwidth]{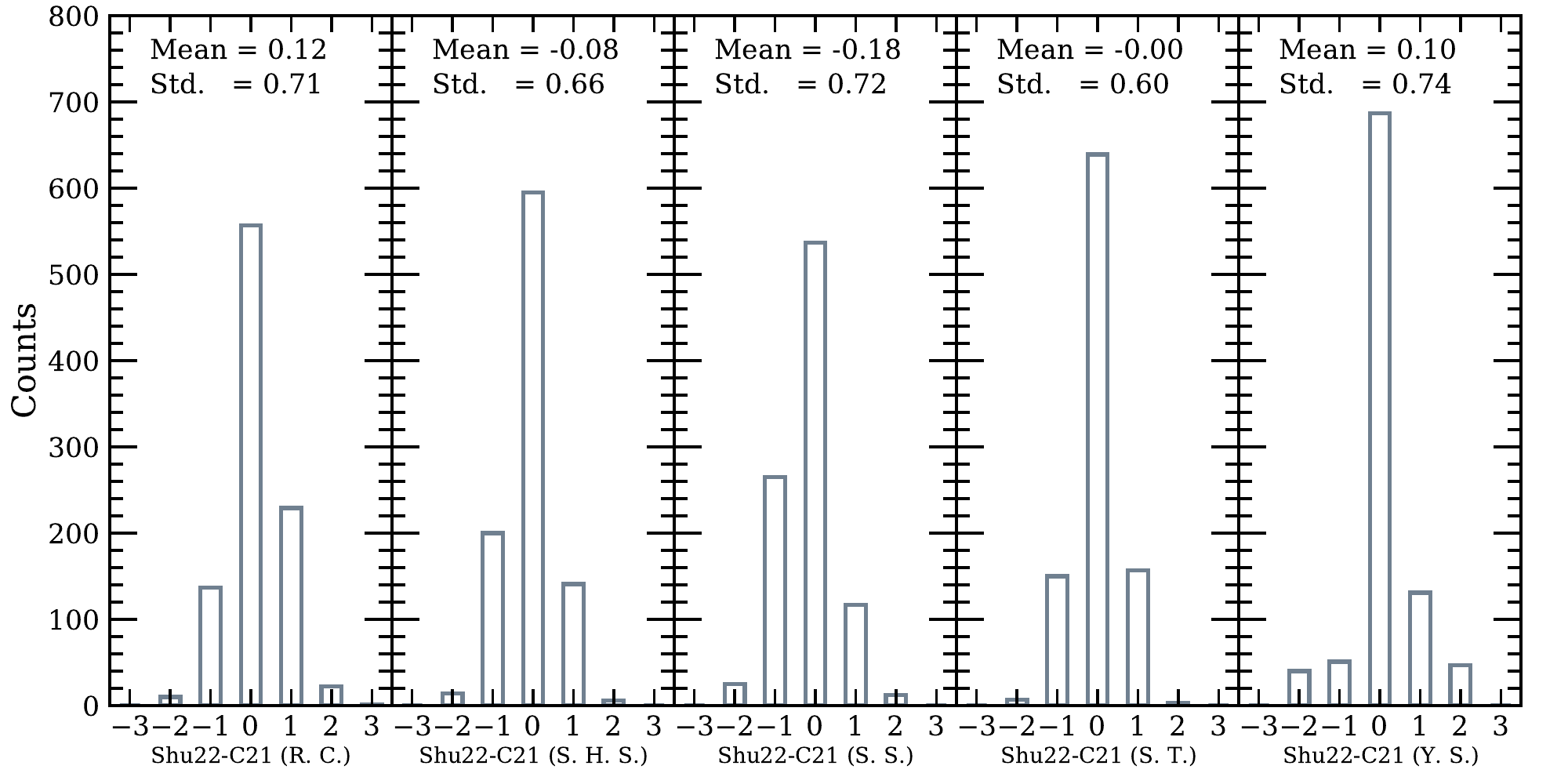}
    \includegraphics[width=0.9\textwidth]{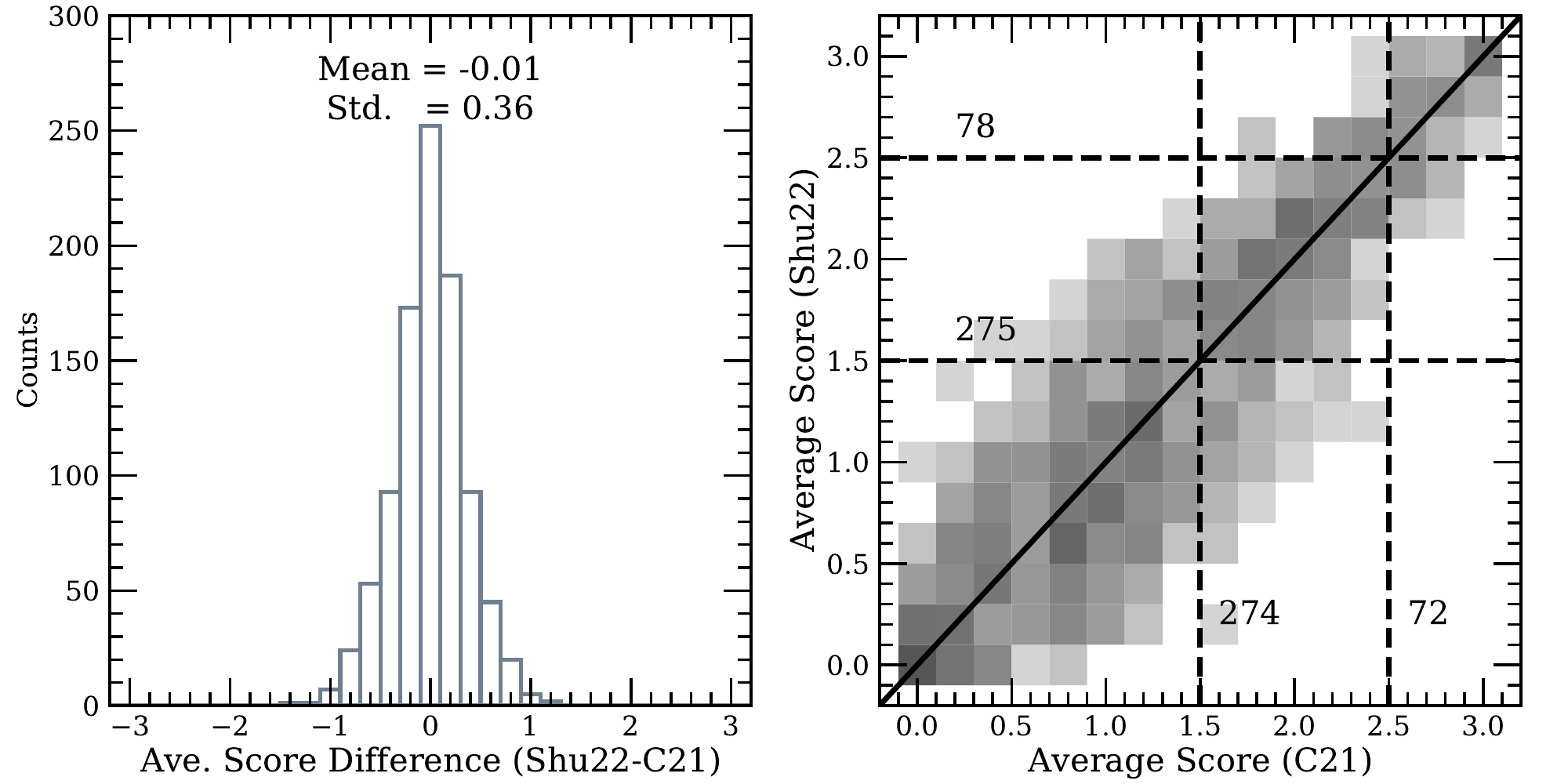}
    \caption{Comparisons on visual-inspection scores for the 956 systems that are in common between this work (i.e. Shu22) and C21. The top row shows the distributions of the difference in scores for the five graders (R. C., S. H. S., S. S., S. T., and Y. S.). The mean and standard deviations of the differences for individual graders are given in each sub-panel. The bottom left panel is the distribution of the difference in the average score, which has a mean of $-0.01$ and standard deviation of $0.36$. The bottom right panel shows the 2D histogram of the average scores in C21 and Shu22. The solid black line is the one-to-one line, and the dashed black lines indicate thresholds that correspond to grade-A or B. According to average scores in C21, 72 are grade As and 274 are grade Bs. According to average scores in this work, 78 are grade As and 275 are grade Bs.}
    \label{fig:comparison_C21}
\end{figure*}

Inevitably, scores from each grader are not all identical. The biases for individual graders range from $-0.18$ to $0.12$, and the typical dispersion is $\sim 0.7$ (Figure~\ref{fig:comparison_C21}). Encouragingly, the average score, which determines the final lens grade, has almost no bias ($-0.01$). Hence, for systems that have different average scores between this work and C21, our recommendation is to adopt the higher values so that a more complete list of candidates can be obtained.

\section{Full lists of grade-A or B lenses}

\begin{table*}[h]
\begin{center} 
\caption{\label{tb:gradeA} List of discovered grade-A strong-lens candidates.}
\begin{tabular}{l c c c c c c c c r}
\hline \hline
Name & R.A. & Decl. & $m_i$ & $z_{\rm d}^{\rm phot}$ & $z_{\rm d}^{\rm spec}$ & $z_{\rm source}^{\rm spec}$ & Classifier & Score & Reference \\
(1) & (2) & (3) & (4) & (5) & (6) & (7) & (8) & (9) & (10) \\
\hline 
HSC\,J000848$+$001550                              & 2.20333      & \phantom{$-$}0.26412      & 18.46  & $0.35   \pm 0.03  $ & 0.397    & ---  & 1      & 2.6  & SuGOHI-5     \\ 
HSC\,J012018$+$001125                              & 20.07557     & \phantom{$-$}0.19048      & 20.14  & $0.63   \pm 0.04  $ & 0.599    & ---  & 1,2    & 2.6  & C21          \\ 
HSC\,J012954$+$003801                              & 22.47583     & \phantom{$-$}0.63363      & 20.27  & $0.67   \pm 0.05  $ & ---  & ---  & 1      & 2.8  & C21          \\ 
HSC\,J015731$-$033057                              & 29.38125     & $-$3.51603      & 20.02  & $0.68   \pm 0.04  $ & 0.621    & ---  & 1,2    & 2.8  & SuGOHI-1     \\ 
HSC\,J015758$-$061426                              & 29.49429     & $-$6.24057      & 18.96  & $0.35   \pm 0.03  $ & ---  & ---  & 1,2    & 2.6  & C21          \\ 
\hline \hline
\end{tabular}
\end{center}
\textsc{      Note.} --- Column 1 is the system name. Columns 2 and 3 are right ascension and declination (J2000) of the lens galaxy. Columns 4 and 5 are the $i-$band CModel magnitude and photometric redshift of the lens galaxy provided by the HSC catalogue. Columns 6 and 7 give the spectroscopic redshifts of the lens and source inferred from auxiliary data. Column 8 indicates the classifier(s) that finds the lens system. Column 9 is the average visual-inspection score of the lens system. Column 10 provides the paper that first discovered the system. Shu22 indicates a completely new discovery. Other relevant references are: Brownstein12: \citet{Brownstein12}; C21: \citet{Canameras21}; Diehl17: \citet{Diehl17}; Huang20: \citet{Huang20}; Huang21: \citet{Huang21}; Jacobs17: \citet{Jacobs17}; Jacobs19b: \citet{Jacobs19b}; Li20: \citet{Li20}; More12: \citet{More12}; More16: \citet{More16}; More17: \citet{More17}; Petrillo19: \citet{Petrillo19}; Ratnatunga95: \citet{Ratnatunga95}; Shu16: \citet{Shu16}; Sonnenfeld13: \citet{Sonnenfeld13}; Stark13: \citet{Stark13}; Stein21: \citet{Stein21}; SuGOHI-1: \citet{sugohi_I}; SuGOHI-2: \citet{sugohi_II}; SuGOHI-4: \citet{sugohi_IV}; SuGOHI-5: \citet{sugohi_V}; SuGOHI-6: \citet{sugohi_VI}. 'Guoyou Sun' corresponds to candidates identified by an amateur astronomer, Guoyou Sun, through visual inspections of HSC cutouts\footnotemark. Systems with $^{\dagger}$ and/or $^{\ddagger}$ are independently discovered by Wong et al. (in prep.) and/or Jaelani et al. (in prep.). The full table is available at the CDS. \\
\end{table*}
\begin{table*}[h]
\begin{center} 
\caption{\label{tb:gradeB} List of discovered grade-B strong-lens candidates.}
\begin{tabular}{l c c c c c c c c r}
\hline \hline
Name & R.A. & Decl. & $m_i$ & $z_{\rm d}^{\rm phot}$ & $z_{\rm d}^{\rm spec}$ & $z_{\rm source}^{\rm spec}$ & Classifier & Score & Reference \\
(1) & (2) & (3) & (4) & (5) & (6) & (7) & (8) & (9) & (10) \\
\hline 
HSC\,J000018$+$001617                              & 0.07884      & \phantom{$-$}0.27158      & 19.91  & $0.63   \pm 0.04  $ & ---  & ---  & 2      & 2.0  & Shu22        \\ 
HSC\,J000020$-$002051                              & 0.08681      & $-$0.34750      & 20.56  & $0.95   \pm 0.05  $ & 0.560    & ---  & 1      & 2.2  & Shu22        \\ 
HSC\,J000106$+$010329$^{\dagger}$                  & 0.27710      & \phantom{$-$}1.05827      & 20.07  & $0.73   \pm 0.03  $ & 0.721    & ---  & 1,2    & 2.0  & Shu22        \\ 
HSC\,J000114$+$001619                              & 0.31063      & \phantom{$-$}0.27214      & 19.41  & $0.67   \pm 0.05  $ & 0.664    & ---  & 1      & 1.6  & Shu22        \\ 
HSC\,J000327$+$021020                              & 0.86261      & \phantom{$-$}2.17248      & 20.18  & $0.72   \pm 0.03  $ & ---  & ---  & 1      & 1.8  & C21          \\ 
\hline \hline
\end{tabular}
\end{center}
\textsc{      Note.} --- The columns are the same as in Table~\ref{tb:gradeA}. The full table is available at the CDS. \\
\end{table*}
\footnotetext{http://sunguoyou.lamost.org/glc.html.}

\begin{figure*}[htbp]
    \centering
    \includegraphics[width=0.96\textwidth]{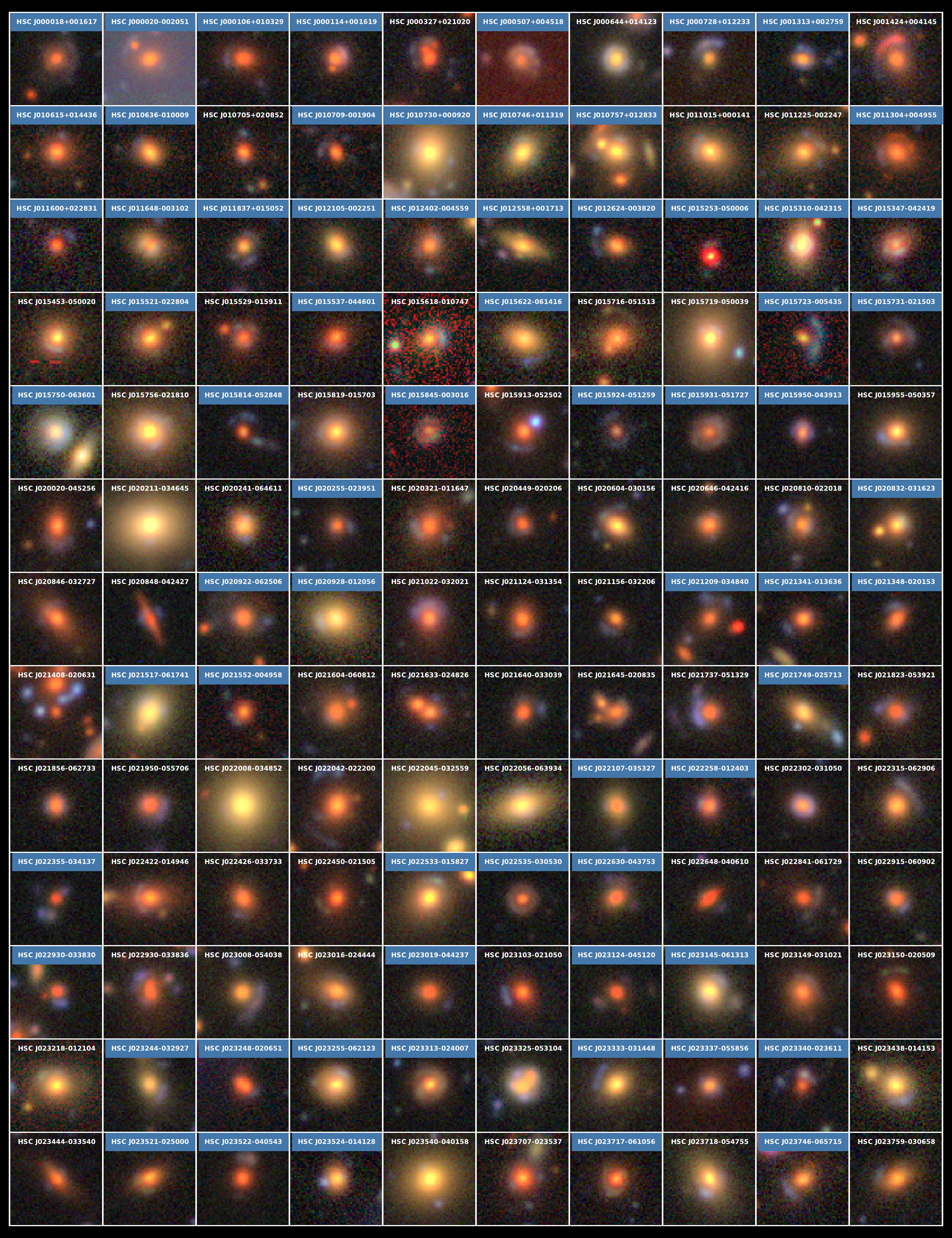}
    \caption{Colour composite images ($10^{\prime \prime} \times 10^{\prime \prime}$) of the 630 grade-B strong-lens candidates discovered by this work. Candidates with blue background beneath the system name are new discoveries.}
    \label{fig:gradeB}
\end{figure*}
\begin{figure*}[htbp]
    \ContinuedFloat
    \centering
    \includegraphics[width=0.96\textwidth]{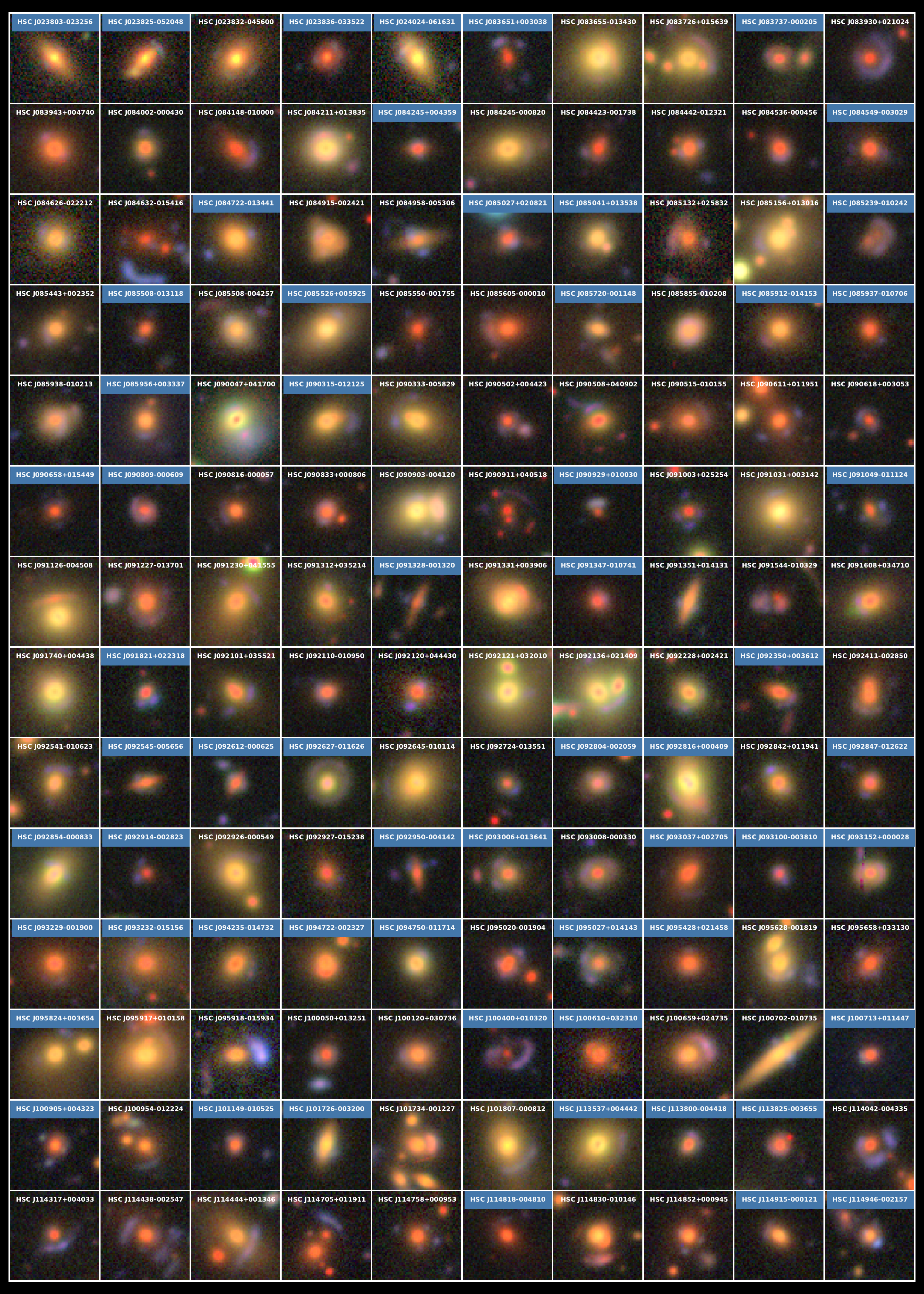}
    \caption{continued.}
\end{figure*}
\begin{figure*}[htbp]
    \ContinuedFloat
    \centering
    \includegraphics[width=0.96\textwidth]{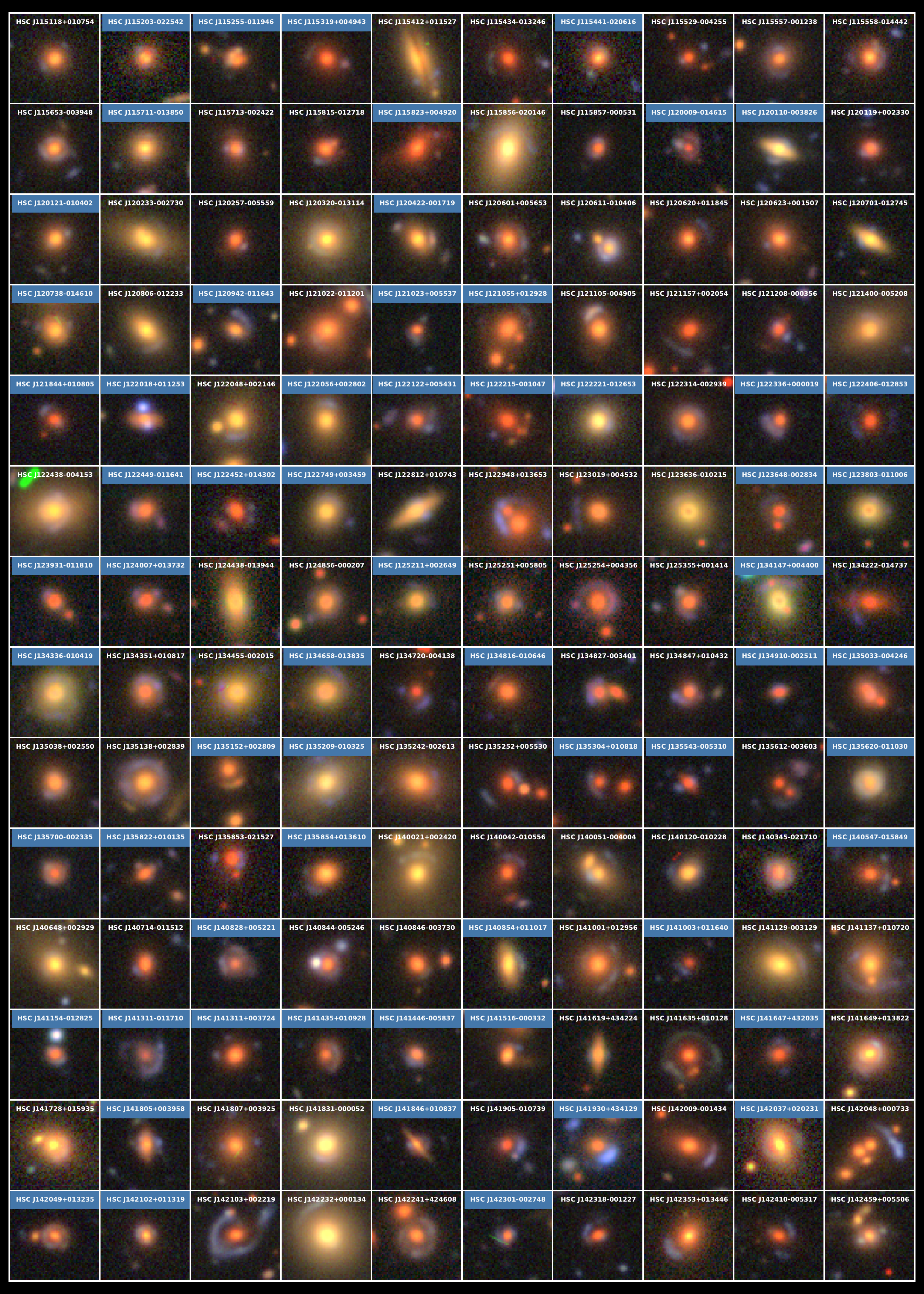}
    \caption{continued.}
\end{figure*}
\begin{figure*}[htbp]
    \ContinuedFloat
    \centering
    \includegraphics[width=0.96\textwidth]{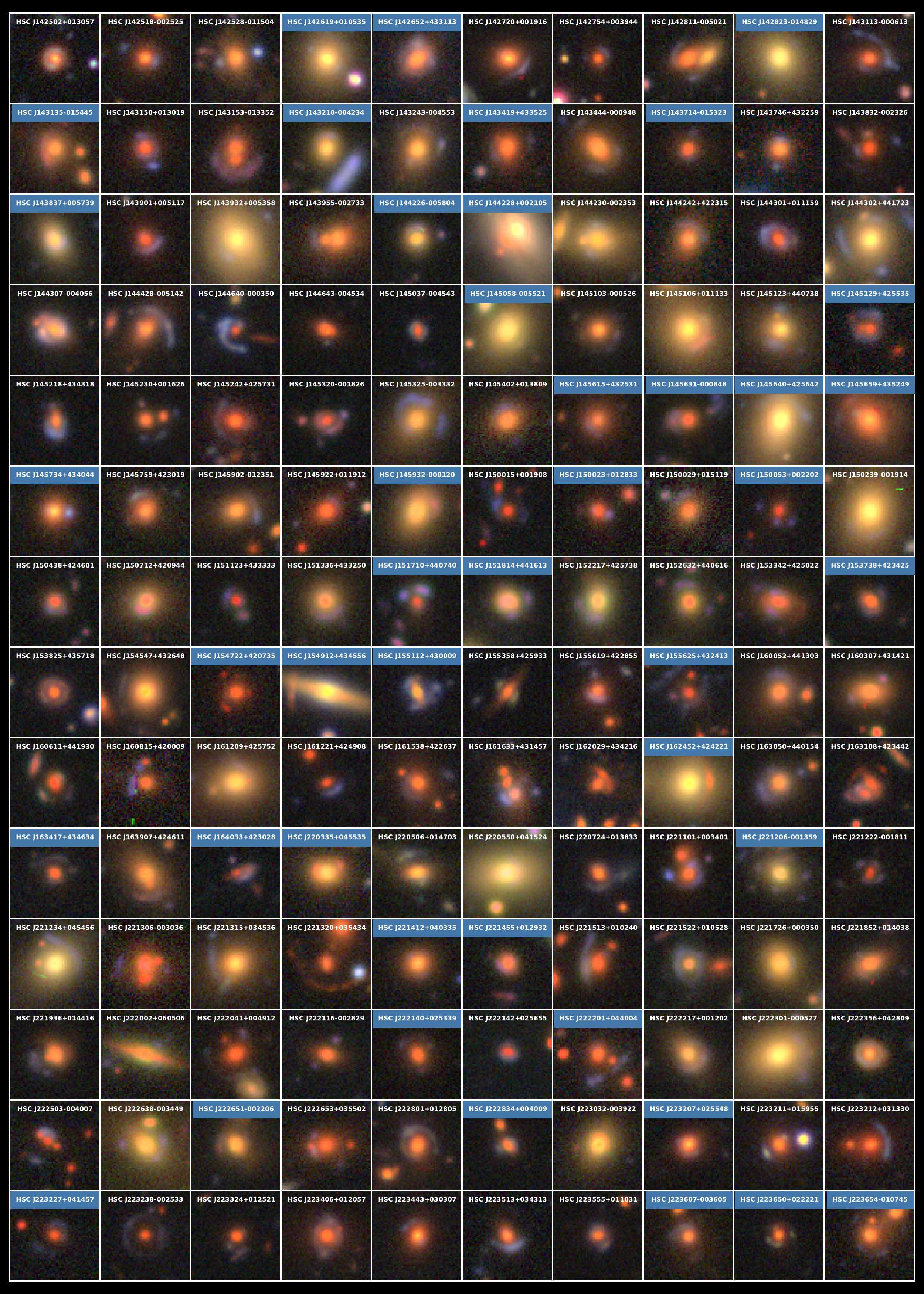}
    \caption{continued.}
\end{figure*}
\begin{figure*}[htbp]
    \ContinuedFloat
    \centering
    \includegraphics[width=0.96\textwidth]{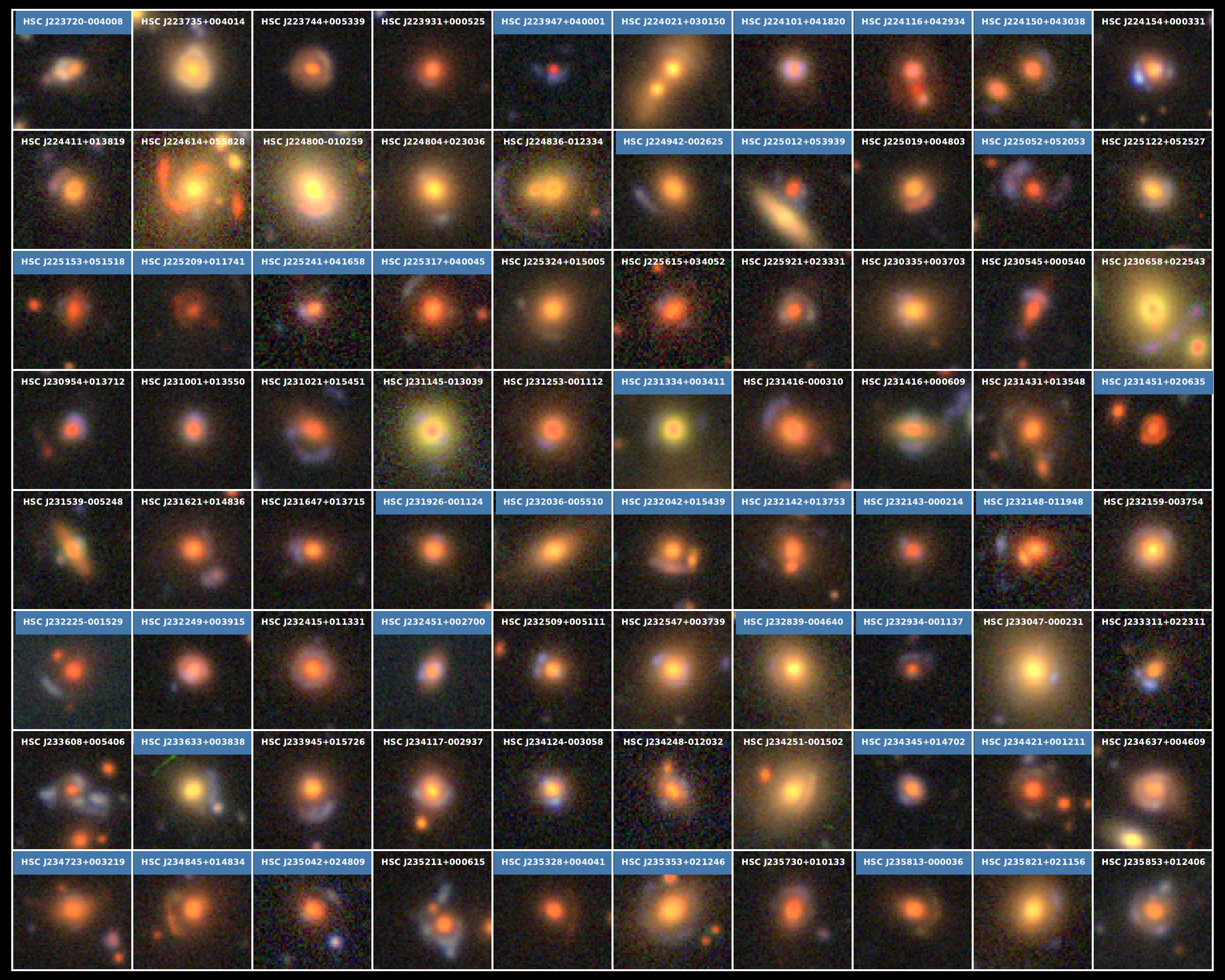}
    \caption{continued.}
\end{figure*}

\end{appendix}

\end{document}